\documentclass[a4paper,11pt]{article}
\usepackage[skins]{tcolorbox}
\usepackage{mathtools,slashed}
\usepackage[T1]{fontenc}
\usepackage{slashed,verbatim,subfigure}
\usepackage[numbers,sort&compress]{natbib}
\usepackage{amsmath}
\usepackage{mathrsfs}
\usepackage{amsbsy}
\usepackage{amssymb}
\usepackage{appendix}
\usepackage{graphicx}
\usepackage[final]{pdfpages}
\usepackage{dcolumn}
\usepackage{bm}
\usepackage{cancel}
\usepackage[colorlinks=true,linktocpage=true,
linkcolor=blue,citecolor=blue]{hyperref}
\usepackage{url}
\definecolor{db5}{cmyk}{0.5,0.5,0,0.5}
\definecolor{mauve}{cmyk}{0.3,0.7,0.1,0.3}
\definecolor{palemauve}{cmyk}{0.3,0.7,0.1,0.0}
\definecolor{pb}{cmyk}{0.4,0.1,0,0.1}
\definecolor{pgreen}{cmyk}{0.4,0.0,0.3,0.0}
\definecolor{pink}{cmyk}{0.0,0.5,0.3,0.0}
\newcommand{\bs}{\begin{slide}}
\newcommand{\es}{\end{slide}}
\newcommand{\tcb}{\textcolor {blue}}
\usepackage[a4paper]{geometry}
\catcode`@11
\def\seceqaa{\@addtoreset{equation}{section}
	\def\theequation{A\arabic{equation}}}
\def\seceqbb{\@addtoreset{equation}{section}
	\def\theequation{B\arabic{equation}}}
\def\seceqcc{\@addtoreset{equation}{section}
	\def\theequation{C\arabic{equation}}}
\def\seceqdd{\@addtoreset{equation}{section}
	\def\theequation{D\arabic{equation}}}
\def\seceqee{\@addtoreset{equation}{section}
	\def\theequation{E\arabic{equation}}}
\def\seceqff{\@addtoreset{equation}{section}
	\def\theequation{F\arabic{equation}}}	
\def\seceqgg{\@addtoreset{equation}{section}
	\def\theequation{G\arabic{equation}}}
\def\seceqhh{\@addtoreset{equation}{section}
	\def\theequation{H\arabic{equation}}}
\def\seceqii{\@addtoreset{equation}{section}
	\def\theequation{H\arabic{equation}}}	
\def\seceqjj{\@addtoreset{equation}{section}
	\def\theequation{H\arabic{equation}}}
\catcode`@11
\newcommand{\be}{\begin{eqnarray}}
\newcommand{\ee}{\end{eqnarray}}
\usepackage{authblk}
\title{QCD-Compatible Supermassive Inert Top-Down Holographic Mesinos at Intermediate Coupling}
\author[,1]{Aalok Misra\footnote{email- aalok.misra@ph.iitr.ac.in}}
\author[,1,2]{Gopal Yadav\footnote{email- gopalyadav@cmi.ac.in, gyadav@ph.iitr.ac.in}\vspace{0.1in}}
\affil[1]{Department of Physics, Indian Institute of Technology Roorkee, Roorkee 247667, Uttarakhand, India}
\affil[2]{Chennai Mathematical Institute,
SIPCOT IT Park, Siruseri 603103, India
}
\date{}
\begin{document}
\maketitle
\begin{abstract}
A longstanding problem with the popular Sakai-Sugimoto holographic dual of thermal QCD is that the ``mesinos'', the (non-supersymmetric) fermionic partners of the mesons, are nearly isospectral with mesons and have an unsuppressed mesino-mesino-meson interaction, both being in contradiction with actual QCD. We solve this problem in a UV complete (and {\it different}) type IIA string dual {\it at intermediate coupling} of realistic thermal QCD, in which the mesinos are shown to be much heavier than and non-interacting with mesons (the wave-function/mass/interaction terms receiving no ${\cal M}$-theory ${\cal O}(R^4)$ corrections). In particular we derive a large-$N$ enhancement of the KK mass scale $M_{KK}$ (from $M_{KK}$ to 
$M_{KK}^{\rm eff}\sim N^{1 + \frac{1}{{\cal O}(1)}}M_{KK}$)  arising from the construction of  the type IIA mirror \cite{MQGP} of the type IIB dual \cite{metrics} of thermal QCD-like theories, as well as the generation of a one-parameter family of $M_{KK}$-independent mass scale at ${\cal O}(R^4)$ in the ${\cal M}$-theory uplift \cite{OR4}  wherein the parameter can be made appropriately large. We also show that the mesino-mesino-single-($\rho/\pi$)meson interactions, vanish identically in the aforementioned type IIA holographic dual.
\end{abstract}

\newpage

\tableofcontents

\section{Introduction}
One can construct gauge theories from a stack of $D$-branes and various configurations of the same. In this context, in the spirit of (non-conformal, non-supersymmetric) gauge-gravity duality (inspired by \cite{Maldacena}),  mostly bosonic fluctuations on the world volume of $D$-branes have been considered. The type IIA dual inclusive of the ${\cal O}(R^4)$ corrections  - to explore the finite-$N$-limit/intermediate coupling regime of QCD -  of the type IIB dual \cite{metrics} of thermal QCD-like theories, was worked out in \cite{MQGP} and \cite{OR4}. As a recent example, inclusive of higher derivative corrections to address the finite-$N$/intermediate coupling regime (as worked out in \cite{OR4}), phenomenologically-compatible low energy coupling constants up to NLO in the chiral expansion in $SU(3)$ chiral perturbation theory (in the chiral limit)  were obtained from the DBI action on flavor $D6$-branes in \cite{Vikas+Gopal+Aalok}. Dirac-like action for the supersymmetric partners of mesons, the mesinos, has been obtained from a top-down approach on $Dp$-branes \cite{Fermionic-action}, see \cite{Abidin+Carlson} for the bottom-up approach. However, using the same for the Sakai-Sugimoto type IIA dual \cite{SS} of thermal QCD, it was shown that one runs into a problem - the mesinos and mesons turn out to be approximately isospectral and their interaction is not large-$N$ suppressed \cite{mesino-SS} - both {\it  in contradiction with real QCD}. This serves as the main motivation for this paper - to see if this issue can be resolved in the type IIA mirror \cite{MQGP} at intermediate coupling \cite{OR4} of the non-supersymmetric {\it UV-complete} type IIB dual \cite{metrics} of thermal QCD-like theories. In this paper, we explicitly consider the mesino action on flavor $D6$-branes in the aforementioned type IIA dual. {\it We also see the effect of higher derivative terms on the fermions relevant to holographic thermal QCD in this paper which was missing in} \cite{Fermionic-action}. {\it In short, we will show that the mesinos are supermassive and do not interact with the vector/$\pi$ mesons, which is why we refer to them as W(eakly) I(nteracting) S(upermassive) P(articles), thereby not being in conflict with realistic QCD.}

The following serves as a brief summary of the main results of this paper. 

\begin{itemize}
\item
{\it Supermassive mesinos} (Sec. \ref{Fermion}):
\begin{itemize}
\item
{\it Dirichlet/Neumann boundary condition for the radial profile of the mesino wave function}:
The on-shell DBI Lagrangian density ${\cal L}_{\rm on-shell}^{\rm DBI,\ D6}$ of the type IIA flavor $D6$-branes (corresponding to $i:\Sigma_{(7)}\cong S^1_t\times_w \mathbb{R}^3\times\mathbb{R}_{\geq0}\times_w S^2_{\rm squashed}(a)\hookrightarrow M_{10}$  [the embedding of the flavor $D6$-branes in the ten-dimensional background involving a warped squashed resolved conifold] in the $\psi=2n\pi, n=0, 1, 2$-coordinate patches and for vanishingly small Ouyang embedding parameter in the parent type IIB dual) obtained from the SYZ mirror of the type IIB holographic dual of \cite{metrics}, in the intermediate-$N$ MQGP limit (\ref{MQGP_limit}), can be shown to be vanishingly small. The mesino EOM,
\begin{eqnarray*}
\label{fermion-EOM}
& & \hskip -0.6in {\cal A}\Theta
+ \Biggl[\frac{\Lambda_2\left(\left\{\Gamma^\alpha\right\}, {\cal F}^{\rm IIA}, {\cal A}\right)}{{\cal L}_{\rm on-shell}^{\rm DBI,\ D6}} + \frac{\Lambda_3\left(\left\{\Gamma^\alpha\right\}, {\cal F}^{\rm IIA}\right)}{\left({\cal L}_{\rm on-shell}^{\rm DBI,\ D6}\right)^2}\Biggr]\Gamma^\gamma D_\gamma\Theta =0, 
\end{eqnarray*} 
(where $\gamma\in\left\{t, x^{1, 2, 3}, r, \theta_2,\tilde{y}\right\}$ indexing coordinates of the flavor $D6$ branes' world volume $\Sigma_{(7)}$, and ${\cal A}, {\cal F}^{\rm IIA}$ are defined in (\ref{Cal-A}), (\ref{calFIIA-def}) respectively; $\Lambda_{2, 3}$ can be read off from (\ref{fermion-EOM})) can hence be approximated by a massless Dirac equation on $\Sigma_{(7)}$.
\item
Either by looking at the $SU(3)$ and the ``transverse'' $SU(3)$ structures on $M_6(= S^1_t\times_w {\cal T},\ \times_w$ implying a warped product, $S^1_t$ being the thermal circle and ${\cal T}$ - deformed $T^{1,1}$ - being the base of a warped non-K\"{a}hler squashed resolved conifold)/$\tilde{M}_6$(=non-K\"{a}hler warped squashed resolved conifold), or when considering the embedding of the $D6$-brane world volume $\Sigma_{(7)}\cong S^1_t\times_w (\mathbb{R}^3\times\mathbb{R}_{\geq 0})\times_w S^2_{\rm squashed}$ in $M_{10}$ considered either as $(S^1_t\times_w\mathbb{R}^3)\times_w \tilde{M}_6$ or 
$\mathbb{R}^3\times_w(\mathbb{R}_{\geq 0}\times M_6)$, one is therefore guaranteed the existence of a pair of globally defined spinors. Using the same, and imposing anti-periodic boundary conditions along $S^1_t$, the ansatz (\ref{mode-ansatz-i}) was made for the mesino spinor, and the radial profile functions therein, were solved for. 
\item
For the thermal background (\ref{TypeIIA-from-M-theory-Witten-prescription-T<Tc}) dual to thermal QCD for $T<T_c$, as well as the black-hole background (\ref{TypeIIA-from-M-theory-Witten-prescription-T>Tc}) dual to thermal QCD for $T>T_c$, we found that Dirichlet/Neumann boundary condition at $r=r_0$ (IR cut-off in the thermal background)/$r=r_h$ permitted supermassive mesinos.

\item
{\it Enhacement of mass scale}:
\begin{itemize}
\item
Starting from the $D=11$ supergravity Einstein's field equations in the presence of four-form $G$ fluxes of ${\cal M}$-theory, we explicitly show the generation of an $N$-enhanced ($\equiv ``N$-hanced'') mass scale, thereby providing the mechanism of generation of super-massive mesinos.
\item
Replacing the resolution parameter ``$a$'' of the blown-up $S^2$  by $a(r)$, substituting an ansatz: $a(r) = b + c^{\beta^0}(r-r_0) + \beta  {\cal A}^\beta(r)$ into the Einstein's equations  and estimating $r_0\sim e^{-\kappa_{r_0}N^{1/3}}$ \cite{Bulk-Viscosity-McGill-IIT-Roorkee}, near the $\psi=2n\pi, n=0, 1, 2$-coordinate patches, we therefore see that: 
\begin{eqnarray}
\label{bare}
& & b \sim N^{1+\frac{1}{{\cal O}(1)}} r_0;\ \ \  {\cal A}^\beta(r) = {\cal C} e^{\frac{c_{\rm linear}}{b}r},\ {\cal C}\equiv{\rm constant}.
\end{eqnarray}
\end{itemize}
\end{itemize}
\item
{\it Vanishing mesino-mesino-meson interaction} (Sec. \ref{No-mesino-meson-interaction}):  Considering fluctuations of the vector mesons $A_{\mu\in S^1_t, \mathbb{R}^3, r}\rightarrow A_{\mu, r}^{(0)} + \delta A_{\mu, r}$ (with $A_{\mu=t}^{(0)}$ being the only non-zero background value) in the fermionic flavor $D6$-brane action and retaining terms linear in the same, performing a KK expansion of the field strength fluctuation along with decomposition of the positive-chirality Majorana-Weyl mesino spinor along $M_5(t, x^{1,2,3}, r)$ and $\tilde{M}_5(\theta_{1,2}, \phi_{1,2}, \psi)$, we were able to show that no mesino-mesino-$\rho/\pi$-meson vertex is generated.
\item
{\it Non-renormalization of the mesino wave function and mass} (Sec. \ref{Fermion}, and appendices \ref{EOM-massive-mesinos} and \ref{gy}):    With the aim of studying the effect of ${\cal O}(R^4)$ terms on the fermions relevant to holographic thermal QCD  which was missing in \cite{Fermionic-action}, leads us {\it  to a non-renormalization of the mesino wave function and mass in the sense that both turn out to be independent of the ${\cal O}(R^4)$ terms up to ${\cal }(l_p^6/N^\alpha), \alpha\geq1$} \footnote{In \cite{OR4}, terms up to ${\cal O}\left(\frac{\beta^0}{N}\right)$ and ${\cal O}\left(\frac{\beta}{N^\alpha}\right),\ 0<\alpha<1,\ \beta\sim l_p^6$, were considered.}, $l_p$ being the Planckian length.
\end{itemize}

The paper is organized as follows. In section \ref{BASICS}, we discuss the type IIA string dual construction of thermal QCD-like theories at intermediate coupling. In section \ref{Fermion}, we show that fermionic superpartner of mesons, i.e., mesinos, are superheavy due to the generation of $N$-enhanced mass scale discussed in section \ref{Enhancement-mass-scale}. Section \ref{No-mesino-meson-interaction} provides further evidence of superheavy mesinos because of the absence of mesino-mesino-meson interaction in type IIA string dual. In section \ref{mq_chiral_cond_RG-inv_OR4-ind}, we discuss the non-renormalization of the product of quark mass and quark condensate up to ${\cal O}(R)^4$. Section \ref{U+L_universality} has a discussion of wave-function universality in the context of glueball, meson, and graviton wave-functions. The summary of the paper has been provided in section \ref{summary}.

There are five appendices. Appendix \ref{finite-mu-baryon} contains the discussion of quark chemical potential. Appendix \ref{EOM-massive-mesinos}  consists of quantities appearing in the mesino EOMs of section  \ref{Fermion}. In Appendix \ref{ztilde-const-incl-beta}, we compute the embedding of flavor $D6$-branes in type IIA string theory inclusive of ${\cal O}(R^4)$ corrections. We list the constants appearing in the wave function for the black hole background in appendix \ref{gy}. Finally, we summarize the top-down holographic QCD results obtained by our group in appendix \ref{Applications_MQGP}.

\section{Type IIA String Dual of Thermal QCD-Like Theories Inclusive of ${\cal O}(R^4)$ Corrections}
\label{BASICS}

Thermal QCD-like theories refer to the equivalence class of theories that are IR confining and UV conformal with the ``quarks'' transforming in the fundamental representation of the symmetry groups(color and flavor). The UV-complete type IIB string dual of such large-$N$ thermal QCD-like theories was constructed in \cite{metrics}. The brane picture consists of  $N$ space-time filling $D3$-branes at the tip of a warped resolved conifold, $M$ space-time filling $D5$ branes also at the tip of the conifold as mentioned above wrapping the vanishing squashed $S^2$ and at the North Pole of the resolved squashed $S^2$ of radius $a$ (resolution parameter), and space-time filling $\overline{D5}$-branes also at the tip of the conifold wrapping the abovementioned vanishing squashed $S^2(\theta_1,\phi_1)$ and at the South Pole of the resolved squashed $S^2(\theta_2,\phi_2)$. In addition, there are $N_f$ space-time filling flavor $D7$-branes wrapping the vanishing squashed $S^3(\theta_1,\phi_1,\psi)$ as well as being at the North Pole of the squashed resolved $S^2(\theta_2,\phi_2)$, dipping into the IR up to $|\mu_{\rm Ouyang}|^{\frac{2}{3}},\ |\mu_{\rm Ouyang}|$ being the modulus of the Ouyang embedding parameter in the Ouyang embedding of the flavor $D7$-branes:  
\begin{equation}
\label{Ouyang}
\left(9 a^2 r^4 + r^6\right)^{1/4}e^{\frac{i}{2}(\psi - \phi_1-\phi_2)}\sin\frac{\theta_1}{2} \sin\frac{\theta_2}{2}=\mu_{\rm Ouyang}.
\end{equation}
 An equal number of $\overline{D7}$ wrapping the vanishing squashed $S^3(\theta_1,\phi_1,\psi)$ and at the South Pole of the blown-up squashed $S^2(\theta_2,\phi_2)$, are also present. Equal number of $D5/D7$-branes and $\overline{D5}/\overline{D7}$-branes in the UV ensure UV conformality. The presence of $N_f$ flavor $D7$ and $\overline{D7}$-branes in the UV, implies a flavor gauge group $SU(N_f) \times SU(N_f)$ in the UV which is broken to $SU(N_f)$ due to absence of $\overline{D7}$-branes in the IR \footnote{On the gravity dual side we characterize UV($r>{\cal R}_{D5/\overline{D5}}$) and IR ($r<{\cal R}_{D5/\overline{D5}}$) in term of radial coordinate where ${\cal R}_{D5/\overline{D5}}$ is the boundary between UV and IR, and separation between $D5$ and $\overline{D5}$-branes.} (analog of chiral symmetry breaking in this brane setup). The brane construct in the type IIB dual is summarized in the table \ref{table-B}:
\begin{table}[h]
\begin{center}
\begin{tabular}{|c|c|c|}\hline
&&\\
S. No. & Branes & World Volume \\
&&\\ \hline
&&\\
1. & $N\ D3$ & $\mathbb{R}^{1,3}(t,x^{1,2,3}) \times \{r=0\}$ \\
&&\\  \hline
&&\\
2. & $M\ D5$ & $\mathbb{R}^{1,3}(t,x^{1,2,3}) \times \{r=0\} \times S^2(\theta_1,\phi_1) \times {\rm NP}_{S^2_a(\theta_2,\phi_2)}$ \\
&&\\  \hline
&&\\
3. & $M\ \overline{D5}$ & $\mathbb{R}^{1,3}(t,x^{1,2,3}) \times \{r=0\}  \times S^2(\theta_1,\phi_1) \times {\rm SP}_{S^2_a(\theta_2,\phi_2)}$ \\
&&\\  \hline
&&\\
4. & $N_f\ D7$ & $\mathbb{R}^{1,3}(t,x^{1,2,3}) \times \mathbb{R}_+(r\in[|\mu_{\rm Ouyang}|^{\frac{2}{3}},r_{\rm UV}])  \times S^3(\theta_1,\phi_1,\psi) \times {\rm NP}_{S^2_a(\theta_2,\phi_2)}$ \\
&&\\  \hline
&&\\
5. & $N_f\ \overline{D7}$ & $\mathbb{R}^{1,3}(t,x^{1,2,3}) \times \mathbb{R}_+(r\in[{\cal R}_{D5/\overline{D5}}-\epsilon,r_{\rm UV}]) \times S^3(\theta_1,\phi_1,\psi) \times {\rm SP}_{S^2_a(\theta_2,\phi_2)}$ \\
&&\\  \hline
\end{tabular}
\end{center}
\caption{The Type IIB Brane Construct of \cite{metrics} (NP and SP respectively denote the North Pole and South Pole of the blown-up $S^2$).}
\label{table-B}
\end{table}

 IR confinement in the gravity dual is affected by deforming the vanishing squashed $S^3$ in the conifold. Since we are interested in finite temperature QCD, the same is effected via the black hole ($T>T_c$) and thermal ($T<T_c$) backgrounds on the gravity dual side. Due to finite temperature and finite separation of $D5$ and $\overline{D5}$-branes on the brane side, the conifold further needs also to possess an $S^2$-blow-up/resolution (with radius/resolution parameter $a$). Additionally, the ten-dimensional warp factor and fluxes include the effect of back-reaction. Therefore, we conclude that string dual of thermal QCD-like theories in the large-$N$ limit involves a warped resolved deformed conifold. The additional advantage of the type IIB dual of \cite{metrics} is that in the IR, at the end of a Seiberg-like duality cascade, the number of colors $N_c$ gets identified with $M$, which in the intermediate-$N$ MQGP limit \cite{MQGP}, \cite{ACMS} 
\begin{equation}
\label{MQGP_limit}
g_s\sim\frac{1}{{\cal O}(1)}, M, N_f \equiv {\cal O}(1), N>1, \frac{g_s M^2}{N}\ll1, \frac{\left(g_s M^2\right)\left(g_s N_f\right)}{N}\ll1,
\end{equation}
 can be tuned to equal 3; given that one is working in the vanishing-Ouyang-modulus limit ($|\mu_{\rm Ouyang}|\ll1$ in (\ref{Ouyang})) of the embedding of the flavor $D7$-branes, $N_f$ can be set to either 2 or 3 corresponding to the lightest quark flavors \cite{Vikas+Gopal+Aalok}.
 
Now, to explore the intermediate coupling regime, the ${\cal O}(R^4)$ terms in eleven-dimensional supergravity action were considered in \cite{OR4}. ${\cal M}$-theory uplift was obtained in two steps: the type IIA Strominger-Yau-Zaslow (SYZ) mirror of type IIB setup was first obtained, and then the former was uplifted to ${\cal M}$-theory. To obtain type IIA SYZ mirror of type IIB setup, a triple T-duality was performed along a local special Lagrangian (sLag) $T^3(x,y,z)$ where $(x,y,z)$ are the toroidal analogs of $(\phi_1,\phi_2,\psi)$ - which could be identified with the $T^2$-invariant sLag of \cite{M.Ionel and M.Min-OO (2008)} - in the large-complex structure limit effected by making the base ${\cal B}(r,\theta_1,\theta_2)$ (of a $T^3(\phi_1,\phi_2,\psi)$-fibration over ${\cal B}(r,\theta_1,\theta_2)$) large \cite{MQGP}, \cite{NPB}. Hence, all the color and flavor $D$-branes get T-dualized to color and flavor $D6$-branes. The ${\cal M}$-theory uplift metric \cite{MQGP}, \cite{OR4} (finite-but-large-$N$/intermediate coupling) of \cite{metrics} (UV-complete type IIB holographic dual of large-$N$ thermal QCD-like theories) is expressed in the following form:
\begin{eqnarray}
\label{TypeIIA-from-M-theory-Witten-prescription-T>Tc}
\hskip -0.1in ds_{11}^2 & = & e^{-\frac{2\phi^{\rm IIA}}{3}}\Biggl[\frac{1}{\sqrt{h(r,\theta_{1,2})}}\left(-g(r) dt^2 + \left(dx^1\right)^2 +  \left(dx^2\right)^2 +\left(dx^3\right)^2 \right)
\nonumber\\
& & \hskip -0.1in+ \sqrt{h(r,\theta_{1,2})}\left(\frac{dr^2}{g(r)} + ds^2_{\rm IIA}(r,\theta_{1,2},\phi_{1,2},\psi)\right)
\Biggr] + e^{\frac{4\phi^{\rm IIA}}{3}}\left(dx^{11} + A_{\rm IIA}^{F_1^{\rm IIB} + F_3^{\rm IIB} + F_5^{\rm IIB}}\right)^2,
\end{eqnarray}
where the type IIA RR 1-forms, $A_{\rm IIA}^{F^{\rm IIB}_{i=1,3,5}}$ are obtained from type IIB $F_{1,3,5}^{\rm IIB}$ fluxes via the SYZ mirror of type IIB string dual \cite{metrics}, $g(r) = 1 - \frac{r_h^4}{r^4}$, and $\phi^{\rm IIA}$ is the type IIA dilaton profile. For low temperatures, i.e., $T<T_c$,  the thermal gravitational dual is given by:
\begin{eqnarray}
\label{TypeIIA-from-M-theory-Witten-prescription-T<Tc}
\hskip -0.1in ds_{11}^2 & = & e^{-\frac{2\phi^{\rm IIA}}{3}}\Biggl[\frac{1}{\sqrt{h(r,\theta_{1,2})}}\left(-dt^2 + \left(dx^1\right)^2 +  \left(dx^2\right)^2 + \tilde{g}(r)\left(dx^3\right)^2 \right)
\nonumber\\
& & \hskip -0.1in+ \sqrt{h(r,\theta_{1,2})}\left(\frac{dr^2}{\tilde{g}(r)} + ds^2_{\rm IIA}(r,\theta_{1,2},\phi_{1,2},\psi)\right)
\Biggr] + e^{\frac{4\phi^{\rm IIA}}{3}}\left(dx^{11} + A_{\rm IIA}^{F_1^{\rm IIB} + F_3^{\rm IIB} + F_5^{\rm IIB}}\right)^2,
\end{eqnarray}
where $\tilde{g}(r) = 1 - \frac{r_0^4}{r^4}$. One notes that $t\rightarrow x^3,\ x^3\rightarrow t$ in (\ref{TypeIIA-from-M-theory-Witten-prescription-T>Tc}) followed by a Double Wick rotation in the new $x^3, t$ coordinates obtains (\ref{TypeIIA-from-M-theory-Witten-prescription-T<Tc}); $h(r,\theta_{1,2})$ is the ten-dimensional warp factor \cite{metrics, MQGP}. This is equivalent to:  $-g_{tt}^{\rm BH}(r_h\rightarrow r_0) = g_{x^3x^3}\ ^{\rm Thermal}(r_0),$ $ g_{x^3x^3}^{\rm BH}(r_h\rightarrow r_0) = -g_{tt}\ ^{\rm Themal}(r_0)$ in the results of \cite{VA-Glueball-decay}, \cite{OR4} (see \cite{Kruczenski et al-2003} in the context of Euclidean/black $D4$-branes in type IIA).  In (\ref{TypeIIA-from-M-theory-Witten-prescription-T<Tc}), we will assume the spatial part of the solitonic $M3$ brane (which, locally, could be interpreted as solitonic $M5$-brane wrapped around a homologous sum of $S^2_{\rm squashed}$ \cite{DM-transport-2014}) and their world volume given by $\mathbb{R}^2(x^{1,2})\times S^1(x^3)$ with the period of $S^1(x^3)$ given by a very large: $\frac{2\pi}{M_{\rm KK}}$, where the very small $M_{\rm KK}$ is given by $\frac{2r_0}{ L^2}\left[1 + {\cal O}\left(\frac{g_sM^2}{N}\right)\right]$, $r_0$ being the very small IR cut-off in the thermal background (see also \cite{Armoni et al-2020}) and $L = \left( 4\pi g_s N\right)^{\frac{1}{4}}$. So, $\lim_{M_{\rm KK}\rightarrow0}\mathbb{R}^2(x^{1,2})\times S^1(x^3) = \mathbb{R}^3(x^{1,2,3})$, thereby recovering 4D physics. The working metric for the thermal background corresponding to $T<T_c$ will involve setting $\tilde{g}(r)$ to unity in (\ref{TypeIIA-from-M-theory-Witten-prescription-T<Tc}).

Eleven dimensional supergravity action including ${\cal O}(R^4)$ terms used in \cite{OR4} is:
\begin{eqnarray}
\label{D=11_O(l_p^6)}
& & \hskip -0.8inS = \frac{1}{2\kappa_{11}^2}\int_M\left[  {\cal R} *_{11}1 - \frac{1}{2}G_4\wedge *_{11}G_4 -
\frac{1}{6}C\wedge G\wedge G\right] + \frac{1}{\kappa_{11}^2}\int_{\partial M} d^{10}x \sqrt{h} K \nonumber\\
& & \hskip -0.8in+ \frac{1}{(2\pi)^43^22^{13}}\left(\frac{2\pi^2}{\kappa_{11}^2}\right)^{\frac{1}{3}}\int d^{11}x\sqrt{-g}\left( J_0 - \frac{1}{2}E_8\right) + \left(\frac{2\pi^2}{\kappa_{11}^2}\right)\int C_3\wedge X_8,
\end{eqnarray}
where:
\begin{eqnarray}
\label{J0+E8-definitions}
& & \hskip -0.8inJ_0  =3\cdot 2^8 (R^{HMNK}R_{PMNQ}{R_H}^{RSP}{R^Q}_{RSK}+
{1\over 2} R^{HKMN}R_{PQMN}{R_H}^{RSP}{R^Q}_{RSK}),\nonumber\\
& & \hskip -0.8inE_8  ={ 1\over 3!} \epsilon^{ABCM_1 N_1 \dots M_4 N_4}
\epsilon_{ABCM_1' N_1' \dots M_4' N_4' }{R^{M_1'N_1'}}_{M_1 N_1} \dots
{R^{M_4' N_4'}}_{M_4 N_4},\nonumber\\
& & \hskip -0.8in\kappa_{11}^2 = \frac{(2\pi)^8 l_p^{9}}{2}.
\end{eqnarray}
The equations of motion for metric and three form potential $C$ are:
\begin{eqnarray}
\label{eoms}
& & {\rm EOM}_{\rm MN}:\ R_{MN} - \frac{1}{2}g_{MN}{\cal R} - \frac{1}{12}\left(G_{MPQR}G_N^{\ PQR} - \frac{g_{MN}}{8}G_{PQRS}G^{PQRS} \right)\nonumber\\
 & &  = - \beta\left[\frac{g_{MN}}{2}\left( J_0 - \frac{1}{2}E_8\right) + \frac{\delta}{\delta g^{MN}}\left( J_0 - \frac{1}{2}E_8\right)\right],\nonumber\\
& & d*G = \frac{1}{2} G\wedge G +3^22^{13} \left(2\pi\right)^{4}\beta X_8,\nonumber\\
& &
\end{eqnarray}
where \cite{Becker-sisters-O(R^4)}:
\begin{equation}
\label{beta-def}
\beta \equiv \frac{\left(2\pi^2\right)^{\frac{1}{3}}\left(\kappa_{11}^2\right)^{\frac{2}{3}}}{\left(2\pi\right)^43^22^{12}} \sim l_p^6,
\end{equation}
$R_{MNPQ}, R_{MN}, {\cal R}$  in  (\ref{D=11_O(l_p^6)})/(\ref{eoms}) are elven-dimensional Riemann curvature tensor, Ricci tensor, and the Ricci scalar. To solve (\ref{eoms}), the following ansatz was made:
\begin{eqnarray}
\label{ansaetze}
& & \hskip -0.8ing_{MN} = g_{MN}^{(0)} +\beta g_{MN}^{(1)},\nonumber\\
& & \hskip -0.8inC_{MNP} = C^{(0)}_{MNP} + \beta C_{MNP}^{(1)}.
\end{eqnarray}
EOM for $C_{MNP}$ symbolically can be written as:
\begin{eqnarray}
\label{deltaC=0consistent}
& & \beta \partial\left(\sqrt{-g}\partial C^{(1)}\right) + \beta \partial\left[\left(\sqrt{-g}\right)^{(1)}\partial C^{(0)}\right] + \beta\epsilon_{11}\partial C^{(0)} \partial C^{(1)} = {\cal O}(\beta^2) \sim 0 [{\rm up\ to}\ {\cal O}(\beta)].
\nonumber\\
& & \end{eqnarray}
It was shown in \cite{OR4}, that, $C^{(1)}_{MNP}=0$ up to ${\cal O}(\beta)$. Therefore only the metric receives ${\cal O}(R^4)$ corrections defined as:
\begin{eqnarray}
\label{fMN-definitions}
\delta g_{MN} =\beta g^{(1)}_{MN} = G_{MN}^{\rm MQGP} f_{MN}(r).
\end{eqnarray}
In general,  the ${\cal M}$ theory metric has the following form including ${\cal O}(R^4)$ corrections: 
\begin{equation}
\label{fMN-def}
G_{MN}^{\cal M} = G_{MN}^{\rm MQGP}\left(1+\beta f_{MN}(r)\right).
\end{equation}
The EOMs for $f_{MN}(r)$ were solved in \cite{OR4}. The type IIA metric inclusive of ${\cal O}(R^4)$ corrections were obtained from the ${\cal M}$-theory metric by descending back to type IIA string theory, which has the following form: 
\begin{equation}
\label{GmnIIAbeta}
 G_{mn}^{\rm IIA}= \sqrt{G_{x^{10}x^{10}}^{\cal M}}G_{mn}^{\rm MQGP}\left(1+\frac{f_{x^{10}x^{10}(r)}}{2}+f_{mn}(r)\right).
\end{equation} 

The type IIB dual of large-$N$ thermal QCD-like theories as constructed in \cite{metrics} and its type IIA mirror as constructed in \cite{MQGP}, \cite{NPB} were successfully used to study a variety of issues in Condensed Matter Physics, lattice/PDG-compatible particle phenomenology, doubly holographic extension and Page curves of associated eternal black holes and $G$/(Almost)Contact(3)Metric structure
classification of underlying six-, seven- and eight-folds in differential geometry (see \ref{Applications_MQGP}).

\section{Supermassive Mesinos in Type IIA String Theory}
\label{Fermion}   
The fermionic sector of type IIA holographic dual of QCD as constructed in \cite{SS} has the following problems. Not only are the mesinos approximately isospectral with the mesons, the single-meson-mesino-mesino interaction terms are not large-$N$ suppressed \cite{mesino-SS} (see also \cite{Nakas+Rigatos} for mesino spectroscopy degenerate with mesons in the context of \cite{SS} and \cite{Kruczenski et al-2003}). Evidently, this is in contradiction with QCD/PDG as no mesino at the EW scale has thus far been observed. What we show in this section is that Dirichlet/Neumann boundary condition at the IR cut-off (for the gravity dual corresponding to $T<T_c$) or the horizon radius (for the gravity dual corresponding to $T>T_c$) is consistent with having a supermassive mesino. Further, we show an $N$-enhancement of the Kaluza-Klein mass scale via an $N$-enhancement of the resolution parameter for the thermal background ($T<T_c$), hence providing the mechanism of generation of the aforementioned supermassive mesino. Even though we have not been able to provide in \ref{Fermion} an analog of the $N$-enhancement of the resolution parameter (that was seen in the thermal background corresponding to $T<T_c$) for the black hole background corresponding to $T>T_c$, the following should be noted. In \ref{Fermion}, what we were able to show for the gravity duals of both the low and high-temperature QCD-like theories is that Dirichlet/Neumann boundary condition at the IR cut-off, horizon radius respectively in the gravity duals for $T<T_c,\ T>T_c$, do not fix the mesino mass. We can hence take the same to be large, and via the aforementioned $N$-enhancement of the resolution parameter in the former, we had explicitly shown the mechanism of obtaining supermassive mesinos in the thermal background. Given that we were able to show the vanishing of meson-mesino-mesino interaction in \ref{No-mesino-meson-interaction}, even if the mesinos were of the EW scale, there still will be no contradiction with real QCD. \par
   The DBI action for the fermions on flavor D6-branes has the following structure \cite{Fermionic-action}:
\begin{eqnarray}
\label{DBI-F}
S_{D_6}^{f}=\frac{T_{D_6}}{2} \int d^7\xi e^{-\Phi^{\rm IIA}}\sqrt{-{\rm det}(i^*g^{\rm IIA}+\mathcal{F}^{\rm IIA})} \  \overline{\Theta}\left(1-\Gamma_{D_6}\right)\left(\Gamma^\alpha D_\alpha-\Delta+L_{D_6}\right)\Theta,
\end{eqnarray}
where $\Phi^{\rm IIA}$ is the type IIA dilaton. We can define: 
\begin{equation}
\label{calFIIA-def}
{\mathcal{F}^{\rm IIA}_{\alpha_{1} \alpha_{2}}} = i^*B_{\alpha_{1} \alpha_{2}}^{\rm IIA}+F_{\alpha_{1} \alpha_{2}}^{\rm IIA},
\end{equation}
 such that $B_{\alpha_{1} \alpha_{2}}^{\rm IIA}$ and $F_{\alpha_{1} \alpha_{2}}^{\rm IIA}$ are NS-NS B field and gauge field restricted to the world volume of $D6$-branes. Further, $\Gamma_{D_6}$ and $L_{D_6}$ appearing in (\ref{DBI-F}) are defined as \footnote{Indices, $m,n,p$ correspond to type IIA bulk indices and $\alpha_{i},\beta_{i},\gamma$ etc. correspond to indices on world-volume of flavor $D6$-branes.}:
\begin{eqnarray}
\label{G-L}
& & \Gamma_{D_6}=\sum_{q+r=3}\frac{(-)^{r+1}\left(\Gamma_{10}\right)^{r+1} \epsilon^{\alpha_1 ....\alpha_{2q}\beta_1.....\beta_{2r+1}}}{q! (2r+1)!2^q\sqrt{-{\rm det}(i^*g^{\rm IIA}+\mathcal{F}^{\rm IIA})}} {\mathcal{F}^{\rm IIA}_{\alpha_{1} \alpha_{2}}} .....\mathcal{F}^{\rm IIA}_{\alpha_{2q-1} \alpha_{2q}} \Gamma_{\beta_1.....\beta_{2r+1}},\nonumber\\
& & \Delta= \Delta^{(1)}+ \Delta^{(2)},
\nonumber\\
& &L_{D_6}=\sum_{q\geq1,q+r=3}\frac{(-)^{r+1}\left(\Gamma_{10}\right)^{r+1} \epsilon^{\alpha_1 ....\alpha_{2q}\beta_1.....\beta_{2r+1}}}{q! (2r+1)!2^q\sqrt{-{\rm det}(i^*g^{\rm IIA}+\mathcal{F}^{\rm IIA})}} {\mathcal{F}_{\alpha_{1} \alpha_{2}}} .....\mathcal{F}_{\alpha_{2q-1} \alpha_{2q}} \Gamma_{\beta_1.....\beta_{2r+1}}^{\ \ \ \ \ \ \ \ \ \ \gamma} D_\gamma,
\end{eqnarray}
where $D_m=D_m^{(0)}+W_m$, and
\begin{eqnarray}
\label{D-Delta}
& & 
D_m^{(0)}= \nabla_m +\frac{1}{4.2!} H_{mnp}\Gamma^{np}\Gamma_{(10)}, \nonumber\\
& & W_m=-\frac{1}{8}e^{\Phi^{\rm IIA}}\left(\frac{1}{2}F_{np}\Gamma^{np} \Gamma_{(10)}+\frac{1}{4!}F_{npqr}\Gamma^{npqr} \right)\Gamma_m ,
\nonumber\\
& & 
\Delta^{(1)}=\frac{1}{2}\left(\Gamma^m \partial_m \Phi^{\rm IIA} +\frac{1}{2.3! }H_{mnp}\Gamma^{mnp} \Gamma_{(10)}\right), \nonumber\\
& & \Delta^{(2)}=\frac{1}{8}e^{\Phi^{\rm IIA}}\left(\frac{3}{2!}F_{mn}\Gamma^{mn} \Gamma_{(10)}-\frac{1}{4!}F_{mnpq}\Gamma^{mnpq} \right), 
\end{eqnarray}
where covariant derivative is defined as: $\nabla_m=\partial_m + \frac{1}{4} \Omega_m^{\underline{np}} \Gamma_{\underline{np}}$. $F_{mn}$ and $F_{mnpq}$ are field strength tensors corresponding to type IIA $A_n$ and $A_{npq}$, and $H_{mnp}=\partial_{[m} B_{np]}$. For flavor $D6$-branes in type IIA string theory
\begin{eqnarray}
\label{LD6-GammaD6-IIA}
& &
\Gamma_{\rm D6} = \frac{\epsilon^{\beta_1.....\beta_7}\Gamma_{\beta_1.....\beta_7}}{\sqrt{-{\rm det}(i^*g^{\rm IIA}+\mathcal{F}^{\rm IIA})}} -\frac{\Gamma_{(10)}\left( \epsilon^{\alpha_1 \alpha_2 \beta_1......\beta_5} {\cal F}_{\alpha_1 \alpha_2}^{\rm IIA} \Gamma_{\beta_1......\beta_5} \right)}{5! \sqrt{-{\rm det}(i^*g^{\rm IIA}+\mathcal{F}^{\rm IIA})}} \nonumber\\
& & + \frac{ \epsilon^{\alpha_1...\alpha_4 \beta_1 \beta_2 \beta_3} \mathcal{F}_{\alpha_1 \alpha_2}^{\rm IIA}\mathcal{F}_{\alpha_3 \alpha_4}^{\rm IIA}\Gamma_{\beta_1 \beta_2 \beta_3} }{48 \sqrt{-{\rm det}(i^*g^{\rm IIA}+\mathcal{F}^{\rm IIA})}} - \frac{\Gamma_{(10)}\left( \epsilon^{\alpha_1....\alpha_6 \beta_1} {\cal F}_{\alpha_1 \alpha_2}^{\rm IIA}{\cal F}_{\alpha_3 \alpha_4}^{\rm IIA}{\cal F}_{\alpha_5 \alpha_6}^{\rm IIA} \Gamma_{\beta_1} \right)}{48 \sqrt{-{\rm det}(i^*g^{\rm IIA}+\mathcal{F}^{\rm IIA})}}, \nonumber\\
& & L_{\rm D6}=-\frac{\Gamma_{(10)}\left( \epsilon^{\alpha_1 \alpha_2 \beta_1......\beta_5} {\cal F}_{\alpha_1 \alpha_2}^{\rm IIA} \Gamma_{\beta_1...\beta_5}^{\ \ \ \ \ \ \gamma}D_\gamma \right)}{240 \sqrt{-{\rm det}(i^*g^{\rm IIA}+\mathcal{F}^{\rm IIA})}}+ \frac{ \epsilon^{\alpha_1...\alpha_4 \beta_1 \beta_2 \beta_3} \mathcal{F}_{\alpha_1 \alpha_2}^{\rm IIA}\mathcal{F}_{\alpha_3 \alpha_4}^{\rm IIA}\Gamma_{\beta_1 \beta_2 \beta_3}^{\ \ \ \ \ \ \gamma}D_\gamma }{48 \sqrt{-{\rm det}(i^*g^{\rm IIA}+\mathcal{F}^{\rm IIA})}}. \nonumber\\
\end{eqnarray}

The Dirac equation for the DBI action for the fermions on flavor $D6$-branes appearing in type IIA string dual of thermal QCD-like theories turns out to be:
{\footnotesize
\begin{eqnarray}
& & \hskip -0.2in \Biggl[{\cal A}-\frac{\epsilon^{\alpha_1 \alpha_{2}\beta_1.....\beta_{5}}\mathcal{F}^{\rm IIA}_{\alpha_{1} \alpha_{2}} \Gamma_{\beta_1.....\beta_{5}}^{\ \ \ \ \ \ \ \ \gamma} D_\gamma\Gamma_{(10)}}{240\sqrt{-{\rm det}(i^*g^{\rm IIA}+\mathcal{F}^{\rm IIA})}}  -  \frac{\epsilon^{\beta_1.....\beta_{7}}\Gamma_{\beta_1.....\beta_{7}}{\cal A}}{7!\sqrt{-{\rm det}(i^*g^{\rm IIA}+\mathcal{F}^{\rm IIA})}} + \tcb{\frac{4\Gamma^{\beta_1.....\beta_{7}}\mathcal{F}^{\rm IIA}_{\beta_{6} \beta_{7}}\Gamma_{\beta_1.....\beta_{5}}^{\ \ \ \ \ \ \ \ \gamma} D_\gamma\Gamma_{(10)}}{7!\left(-{\rm det}(i^*g^{\rm IIA}+\mathcal{F}^{\rm IIA})\right)}} \nonumber\\
& &+\frac{\epsilon^{\alpha_1 \alpha_{2}\beta_1.....\beta_{5}}\mathcal{F}^{\rm IIA}_{\alpha_{1} \alpha_{2}} \Gamma_{\beta_1.....\beta_{5}} \Gamma_{(10)} {\cal A}}{5!\sqrt{-{\rm det}(i^*g^{\rm IIA}+\mathcal{F}^{\rm IIA})}} + \tcb{\frac{7!\mathcal{F}^{\rm IIA}_{\alpha_{1} \alpha_{2}}\mathcal{F}_{\rm IIA} ^{\alpha_{1} \alpha_{2}}\Gamma^{\beta_1.....\beta_{5}}\Gamma_{\beta_1.....\beta_{5}}^{\ \ \ \ \ \ \ \ \gamma} D_\gamma\Gamma_{(10)}}{5!240\left(-{\rm det}(i^*g^{\rm IIA}+\mathcal{F}^{\rm IIA})\right)}}
+\frac{ \epsilon^{\alpha_1...\alpha_4 \beta_1 \beta_2 \beta_3} \mathcal{F}_{\alpha_1 \alpha_2}^{\rm IIA}\mathcal{F}_{\alpha_3 \alpha_4}^{\rm IIA}\Gamma_{\beta_1 \beta_2 \beta_3} {\cal A}}{48 \sqrt{-{\rm det}(i^*g^{\rm IIA}+\mathcal{F}^{\rm IIA})}} \nonumber\\ & &
-{\tcb{\frac{\Gamma_{(10)} {\cal F}_{\rm IIA}^2 {\cal F}_{\beta_4 \beta_5}^{\rm IIA} \Gamma_{\beta_1 \beta_2 \beta_3}\Gamma^{\beta_1.....\beta_{5}\gamma} D_\gamma}{480 \left(-{\rm det}(i^*g^{\rm IIA}+\mathcal{F}^{\rm IIA})\right)}} }+ \tcb{\frac{7!{\cal F}_{\rm IIA}^4\Gamma_{\beta_1 \beta_2 \beta_3}\Gamma^{\beta_1\beta_2 \beta_3\gamma} D_\gamma}{48 \left(-{\rm det}(i^*g^{\rm IIA}+\mathcal{F}^{\rm IIA})\right)}}-\frac{\Gamma_{(10)}\left( \epsilon^{\alpha_1....\alpha_6 \beta_1} {\cal F}_{\alpha_1 \alpha_2}^{\rm IIA}{\cal F}_{\alpha_3 \alpha_4}^{\rm IIA}{\cal F}_{\alpha_5 \alpha_6}^{\rm IIA} \Gamma_{\beta_1} \right) {\cal A}}{48 \sqrt{-{\rm det}(i^*g^{\rm IIA}+\mathcal{F}^{\rm IIA})}} \nonumber
\end{eqnarray}
\begin{eqnarray}
\label{fermion-EOM}
& & \hskip -0.3in +\tcb{\frac{\delta^{[\alpha_3}_{[\beta_2}\delta^{\alpha_4}_{\beta_3}\delta^{\alpha_5}_{\beta_4}\delta^{\alpha_6]}_{\beta_5]} {\cal F}_{\rm IIA}^2{\cal F}_{\alpha_3 \alpha_4}^{\rm IIA} {\cal F}_{\alpha_5 \alpha_6}^{\rm IIA} \Gamma_{\beta_1}\Gamma^{\beta_1...\beta_{5}\gamma} D_\gamma }{5760\left(-{\rm det}(i^*g^{\rm IIA}+\mathcal{F}^{\rm IIA})\right)}}-  \tcb{\frac{ {\cal F}_{\rm IIA}^4 {\cal F}_{\beta_2 \beta_3}^{\rm IIA} \Gamma_{\beta_1}\Gamma^{\beta_1\beta_2\beta_{3}\gamma} D_\gamma }{24\left(-{\rm det}(i^*g^{\rm IIA}+\mathcal{F}^{\rm IIA})\right)}}+ \frac{ \epsilon^{\alpha_1...\alpha_4 \beta_1 \beta_2 \beta_3} \mathcal{F}_{\alpha_1 \alpha_2}^{\rm IIA}\mathcal{F}_{\alpha_3 \alpha_4}^{\rm IIA}\Gamma_{\beta_1 \beta_2 \beta_3}^{\ \ \ \ \ \ \gamma}D_\gamma }{48 \sqrt{-{\rm det}(i^*g^{\rm IIA}+\mathcal{F}^{\rm IIA})}} \nonumber\\
& &  
-\tcb{\frac{\delta_{[\alpha_1}^{[\beta_4}\delta_{\alpha_2}^{\beta_5}\delta_{\alpha_3}^{\beta_6}\delta_{\alpha_4]}^{\beta_7]} {\cal F}_{\rm IIA}^2{\cal F}_{\rm IIA}^{\alpha_1 \alpha_2} {\cal F}_{\rm IIA}^{\alpha_3 \alpha_4} \Gamma^{\beta_1 \beta_2 \beta_3}\Gamma_{\beta_1...\beta_{7}}^{\ \ \ \ \ \ \ \gamma} D_\gamma }{48\left(-{\rm det}(i^*g^{\rm IIA}+\mathcal{F}^{\rm IIA})\right)}}-\tcb{\frac{ {\cal F}_{\rm IIA}^2 {\cal F}_{\rm IIA}^{\beta_4 \beta_5} \Gamma_{\beta_1... \beta_5}\Gamma^{\beta_1\beta_2\beta_{3}\gamma} D_\gamma}{120 \left(-{\rm det}(i^*g^{\rm IIA}+\mathcal{F}^{\rm IIA})\right)}}
\Biggr]\Theta =0,
\end{eqnarray}
}
where
\begin{eqnarray}
\label{Cal-A}
& & {\cal A}=\Gamma^\alpha D_\alpha-\frac{1}{2}\left(\Gamma^m \partial_m \Phi^{\rm IIA} +\frac{1}{12} H_{mnp}\Gamma^{mnp} \Gamma_{(10)}\right)-\frac{1}{8}e^{\Phi^{\rm IIA}} \left(\frac{3}{2}F_{mn}\Gamma^{mn} \Gamma_{(10)}-\frac{1}{4!}F_{mnpq} \Gamma^{mnpq} \right). \nonumber\\
\end{eqnarray}
In (\ref{Cal-A}), $F_{mnpq}$ is the type IIA RR four-form field strength. This in our computation is set to zero as one can show that the same can not be generated by a triple T dual of the RR $F^{IIB}_{1, 3, 5}$ \cite{MQGP}. Type IIA NS-NS $B$ is given by \cite{V+A+K}:
{\footnotesize
\begin{eqnarray}
\label{B-IIA-diag-diag}
& & \hskip -0.4in  B^{\rm IIA}\left(\theta_1=\frac{\alpha_{\theta_1}}{N^{\frac{1}{5}}},\theta_2\sim\frac{\alpha_{\theta_2}}{N^{\frac{3}{10}}}\right) =  d\theta_2\wedge d\tilde{x} \left(-\frac{2 \sqrt[4]{\pi } \sqrt[4]{{g_s}} N^{3/4} \left(3 \sqrt{6} \alpha_{\theta_1}^3-2 \alpha_{\theta_1}^2 \sqrt[5]{N}+2 \alpha_{\theta_2}^2\right)}{27 \alpha_{\theta_1}^4 \alpha_{\theta_2}}\right)\nonumber\\
   & & \hskip -0.4in
+d\theta_2\wedge d\tilde{y} \left(\frac{2 \sqrt[4]{\pi } \sqrt[4]{{g_s}} N^{3/4} \left(3 \sqrt{6} \alpha_{\theta_1}^3-2 \alpha_{\theta_1}^2 \sqrt[5]{N}+2 \alpha_{\theta_2}^2\right)}{27 \alpha_{\theta_1}^4 \alpha_{\theta_2}}\right)  + d\theta_2\wedge d\tilde{z}\left(-\frac{\sqrt[4]{\pi } \alpha_{\theta_2} \sqrt[4]{{g_s}} N^{3/20} \left(2 \left(\sqrt[3]{3}-1\right) \alpha  \sqrt[10]{N}+\sqrt[3]{3}
   \alpha_{\theta_2}\right)}{3^{5/6} \alpha  \sqrt{\alpha_{\theta_2}^2}}\right).\nonumber\\
   & & 
\end{eqnarray}} 
When we restrict to the world-volume of $D6$-branes, then only the non-trivial component that survives will be $B^{\rm IIA}_{\theta_2 \tilde{y}}$. The induced metric on the world volume of $D6$-branes can be obtained from the target space metric as given below:
\begin{eqnarray}
\label{induced-metric-D6}
ds_{\rm D6}^2=ds^2_5+g^{\rm IIA}_{\theta_2 \theta_2} d\theta_2^2+g^{\rm IIA}_{\theta_2 \tilde{y}} d\theta_2 d\tilde{y} + g^{\rm IIA}_{\tilde{y} \tilde{y}} d\tilde{y}^2.
\end{eqnarray}
 Typically, type IIA metric is not diagonal in the basis $(x,y,z)$. Since we need the metric component along $\tilde{y}$-direction therefore, we are writing the metric in diagonal basis in subspace  $(\tilde{x},\tilde{y}, \tilde{z})$ \cite{V+A+K}:
 \begin{eqnarray}
 \label{diagonal-metric}
& &
 ds^2=\frac{2 d\tilde{x}^2 \left(9 \sqrt{2} \sqrt[6]{3} \alpha_{\theta_1} N^{4/5}-2\ 3^{2/3} N\right)}{27 \alpha_{\theta_1}^2 \alpha_{\theta_2}^2}+\frac{2 {d\tilde{y}}^2 \left(2\ 3^{2/3}
   N-9 \sqrt{2} \sqrt[6]{3} \alpha_{\theta_1}  N^{4/5}\right)}{27 \alpha_{\theta_1}^2 \alpha_{\theta_2}^2}\nonumber\\
   & &+\frac{2 d\tilde{z}^2 \left(3^{2/3} \alpha_{\theta_1}^2 N^{3/5}+3^{2/3} \alpha_{\theta_2}^2
   N^{2/5}\right)}{27 \alpha ^2 \alpha_{\theta_2}^2}.
\end{eqnarray}
$ds^2_5$ in (\ref{induced-metric-D6}) is non-compact metric listed along $(t,x^{1,2,3},r)$ subspace, and from (\ref{diagonal-metric}), $g^{\rm IIA}_{\theta_2 \tilde{y}}=0$ and $g^{\rm IIA}_{\tilde{y} \tilde{y}}=\frac{2  \left(2\ 3^{2/3}
   N-9 \sqrt{2} \sqrt[6]{3} \alpha_{\theta_1}  N^{4/5}\right)}{27 \alpha_{\theta_1}^2 \alpha_{\theta_2}^2}$.

Consider the DBI action on the world volume of flavor $D6$-branes: 
\begin{equation}
\label{SDBI}
S_{\rm DBI}^{\rm D6}= -T_{D6}N_f\int_{\Sigma_{(7)}}
\sqrt{-{\rm det}(i^*\left(g^{\rm IIA}+B^{\rm IIA}\right)+F^{\rm IIA})},
\end{equation}
  $i:\Sigma_{(7)}\cong S^1_t\times_w \mathbb{R}^3\times\mathbb{R}_{\geq0}\times_w S^2_{\rm squashed}(a)\hookrightarrow M_{10}$  [the embedding of the flavor $D6$-branes in the ten-dimensional background involving a warped squashed resolved conifold] in the $\psi=2n\pi, n=0, 1, 2$-coordinate patches and vanishingly small Ouyang embedding parameter in the parent type IIB dual. Using the induced metric on the flavor $D6$-branes as given in (\ref{induced-metric-D6}), NS-NS $B^{\rm IIA}$ as given in (\ref{B-IIA-diag-diag}) and turning on a quark chemical potential (by looking at the DBI action in the UV and solving for $A_t(r)$ - see (\ref{At-UV}))  corresponding to $U(1)$ sub-group of $U(N_f)$ with the associated field strength $F_{rt} = A_t'(r)$, the background $A_t(r)$ can be obtained (see appendix \ref{finite-mu-baryon}). In the IR,
 ${\cal L}_{\rm DBI,\ on-shell}^{\rm D6} $, for $N\sim10^2$, can be shown to be infinitesimal.

The coefficient of the most dominant (quadratic) powers of $\frac{1}{{\cal L}_{\rm DBI,\ on-shell}^{\rm D6}}$ in (\ref{fermion-EOM})  marked in blue, is proportional to $\Gamma^\gamma D_\gamma\Theta, \gamma\in\left.\left\{t, x^{1, 2, 3}, r, \theta_2,\tilde{y}\right\}\right|_{\left\{\tilde{x}=0, \tilde{z}={\rm constant}\right\}}$  where  $(\tilde{x}, \tilde{y}, \tilde{z})$ diagonalize $T^3(x, y, z)$ of \ref{BASICS}. One can further show that in the MQGP limit, $E^\gamma_{\underline{a}}\Gamma^{\underline{a}}D_\gamma\Theta\approx0$. The non-K\"{a}hler six-fold $M_6 = S^1_t\times_w {\cal T}$ ($\times_w$ implying a warped product), $S^1_t$ being the thermal circle and ${\cal T}$ - deformed $T^{1,1}$ - being the base of a warped non-K\"{a}hler squashed resolved conifold, was shown to possess an $SU(3)$ structure in \cite{OR4}, with another ``transverse'' $SU(3)$ structure induced from the (Almost) Contact Metric Structure \cite{ACMS} arising from the $G_2$ structure of warped product of the ${\cal M}$-theory circle and $M_6$. Further, the non-K\"{a}hler warped squashed resolved conifold $\tilde{M}_6$ in the type IIA dual also possesses an $SU(3)$ structure \cite{NPB}, \cite{OR4}. Either way, one is therefore guaranteed the existence of a pair of globally defined spinors $\Theta_{1, 0}$ and $\Theta_{2, 0}$ (either by looking at the $SU(3)$ and the ``transverse'' $SU(3)$ structures on $M_6/\tilde{M}_6$ or when considering the embedding of the $D6$-brane world volume $\Sigma_{(7)}\cong S^1_t\times_w (\mathbb{R}^3\times\mathbb{R}_{\geq 0})\times_w S^2_{\rm squashed}$ in $M_{10}$ considered either as $(S^1_t\times_w\mathbb{R}^3)\times_w \tilde{M}_6$ or 
$\mathbb{R}^3\times_w(\mathbb{R}_{\geq 0}\times M_6)$). Making an ansatz:
\begin{eqnarray}
\label{mode-ansatz-i}
& & \Theta_i(x^\mu,y^m) = \Theta_i(t, x^1, r, \theta_2) = \sum_{n:-\infty}^\infty T_n(t) e^{- \sqrt{-1} p x^1}R_{n, i}(r)\left(1 + \beta f_i(\theta_2)\right)\Theta_{i, 0}, i=1, 2,
\end{eqnarray}
$\beta\sim l_p^6$ ($l_p$ being the Planckian length) and assuming $T_n(t) =  e^{i(2n+1)\pi T t}$ (as one imposes anti-periodic boundary conditions on the fermions along the thermal circle thereby breaking supersymmetry \cite{Witten-Hawking-Page-Tc}) implying $\Theta(t+1/T, r) = - \Theta(t,r)$, (and after a double Wick rotation along $t, x^1$, $p^2 = - m_{\rm Mesino}^2$ with $m_{\rm Mesino}$ being the non-supersymmetric mesino mass in the holographic dual of ${\rm QCD}_{\rm Mesino}$) analogous to the relation between the killing spinors $\epsilon_{1,2}$ for a supersymmetric $D6$-brane in flat space: $\epsilon_1 = \Gamma^{\underline{8}\underline{9}\ \underline{10}}\epsilon_2$, we will impose, by hand, and for our non-supersymmetric model: {\footnotesize $\Theta_{1, 0} = \Gamma^{\underline{6}\underline{8}\ \underline{10}}\Theta_{2, 0},$}
in a curved space, where {\footnotesize $\Theta_{1,2}\equiv\frac{1}{2}\left({\bf 1}+/- \Gamma^{(10)}\right)\Theta$, $\Gamma^{(10)} \equiv \prod_{\underline{A}=\underline{0}}^{\underline{9}}\Gamma^{\underline{A}}$; $\underline{A}$}, with $A=1,..., 10$ denoting the ten-dimensional tangent space indices. 

The most dominant spin-connection terms in the IR are contained in $ E^r_{\underline{5}}\Gamma^{\underline{5}}D_r\Theta$, in particular
$\omega_r^{\underline{7}\ \underline{10}}/\omega_r^{\underline{8}\ \underline{10}}$ respectively for the thermal(``TH''), black-hole (``BH'') backgrounds. Consequently, substituting (\ref{mode-ansatz-i}) into $\Theta$'s EOM (details given in this section and \ref{EOM-massive-mesinos}), the same at ${\cal O}(\beta)$ is:
{\footnotesize
\begin{eqnarray}
\label{ang-T_greater_than_Tc}
& &  \Biggl[i(2n+1)\pi T R_{2, n}(r) f_2(\theta_2) + \frac{E^{\theta_2}_{\underline{7}}}{E^t_{\underline{1}}}\Gamma^{\underline{1}\underline{7}}R_{2,n}(r)f_2^\prime(\theta_2)
+ \frac{E^r_{\underline{5}}}{E^t_{\underline{1}}}\Gamma^{\underline{1}\underline{5}}R'_{2, n}(r)f_2(\theta_2) - i p \frac{E^{x^1}_{\underline{2}}}{E^t_{\underline{1}}}\Gamma^{\underline{1}\underline{2}}R_{2, n}(r)f_2(\theta_2)\Biggr]\Theta_{2,0}
\nonumber\\
& & - {\cal J}(r)R_{1, n}(r)f_1(\theta_2)\Gamma^{\underline{15678}}\Theta_{1,0} = 0,
\end{eqnarray}
}
with ${\cal J}\equiv \omega_r^{\underline{7}\ \underline{10}}\frac{E^r_{\underline{5}}}{E^t_{\underline{1}}}$ ($E_{\underline{a}}^M$ being the frames: $E_{\underline{a}}^M g_{MN}E^N_{\underline{b}}=\eta_{\underline{ab}}$) for the TH background; for the BH background, $\Gamma^{\underline{15678}}$ in the second line of (\ref{ang-T_greater_than_Tc}) is to be replaced by $\Gamma^{\underline{156}}$ with ${\cal J}\equiv \omega_r^{\underline{8}\ \underline{10}}\frac{E^r_{\underline{5}}}{E^t_{\underline{1}}}$. Note, we have disregarded all ${\cal O}\left(\frac{\beta}{N^\alpha}\right),\ \alpha\geq1$ terms (see footnote {\bf 1}) and therefore there  are no $\beta$ corrections in $\frac{E^{\theta_2}_{\underline{7}}}{E^t_{\underline{1}}},  \frac{E^r_{\underline{5}}}{E^t_{\underline{1}}},  \frac{E^{x^1}_{\underline{2}}}{E^t_{\underline{1}}}$. One thus sees that the only consistent solution for $f_i(\theta_2)$ is $f_i(\theta_2)=0$ for the TH/BH backgrounds.

Defining $u \equiv \sqrt{r - r_0}$, the EOM for $R_{n,2}(r)$ for the TH type IIA background with $\Gamma^{\underline{15}}\Theta_{2,0} = \Theta_{2,0},\ \Gamma^{\underline{12}}\Theta_{2,0} = \Theta_{2,0},$  can be  recast into a Schr\"{o}dinger-like equation (where, $a_1, b_1, {\cal A}_{\Theta_2}, {\cal B}_{\Theta_2}, {\cal A}_{\Theta_2'}, {\cal B}_{\Theta_2'}$ are defined in (\ref{constants-exps})):
\begin{eqnarray}
\label{EOM-thermal-Schr}
& & \chi_{2, n}''(u) + V(u) \chi_{2, n}(u) = 0,
\end{eqnarray}
where,
\begin{eqnarray}
\label{V-Schroedinger-like}
& & V(u) =  -\frac{3}{4u^2} + \frac{{\cal A}_{{\Theta_2}'}}{a_1 u}  - \frac{{\cal A}_{{\Theta_2}'}}{a_1^2} + {\cal O}(u),
\end{eqnarray}
and,
{\footnotesize
\begin{equation}
\label{EOM-thermal-iii}
 R_{2, n}(u) = \sqrt{u} \left({a_1}+{b_1} u^2\right)^{-\frac{ {\cal B}_{{\Theta_2}'}}{2 {b_1}}} e^{-\frac{{\cal A}_{{\Theta_2}'} \tan
   ^{-1}\left(\frac{\sqrt{{b_1}} u}{\sqrt{{a_1}}}\right)}{\sqrt{{a_1}} \sqrt{{b_1}}}} \chi_{2, n}(u).
\end{equation}
}
The solution of (\ref{EOM-thermal-Schr}) is given by:
{\footnotesize
\begin{eqnarray}
\label{EOM-thermal-iv}
& & \chi_{2, n}(u)=c_{1, n} M_{\frac{1}{2},1}\left(\frac{2{\cal A}_{{\Theta_2}'} u}{{a_1}}\right)+c_{2, n} W_{\frac{1}{2},1}\left(\frac{2 {\cal A}_{{\Theta_2}'}
   u}{{a_1}}\right).
\end{eqnarray}
}
One, therefore obtains:
{\footnotesize
\begin{eqnarray}
\label{EOM-thermal-v}
& & \hskip -0.4in R_{2, n}(r\sim r_0) = \frac{c_{2, n} {a_1}^{\frac{1}{2}-\frac{{\cal B}_{{\Theta_2}'}}{2 {b_1}}}}{\sqrt{2} \sqrt{{\cal A}_{{\Theta_2}'}}}-\frac{(r - r_0)
   {a_1}^{-\frac{{\cal B}_{{\Theta_2}'}}{2 {b_1}}-\frac{3}{2}} \left({a_1} {\cal B}_{{\Theta_2}'} c_{2, n}+{\cal A}_{{\Theta_2}'}^2 (4 c_{2, n}-8 c_{1, n})\right)}{2
   \sqrt{2} \sqrt{{\cal A}_{{\Theta_2}'}}} + {\cal O}\left((r - r_0)^{3/2}\right).
   \nonumber\\ 
\end{eqnarray} 
}
From (\ref{constants-exps}), one sees the absence of ${\cal O}(R^4)$ corrections in (\ref{EOM-thermal-v}) . One also sees from (\ref{EOM-thermal-v}) that one can impose Dirichlet boundary condition at $r=r_0$ (thereby setting $c_2=0$) for all and hence superheavy mesinos ($M_{\rm Mesino}$).

For the BH background assuming $\Gamma^{\underline{15}}\Theta_{2,0} = \Theta_{2,0},\ \Gamma^{\underline{12}}\Theta_{2,0} = \Theta_{2,0},$
implying, $ \Gamma^{\underline{25}}\Theta_{2,0} = \Theta_{2,0},$ in the IR (i.e., near $r=r_h$), redefining $u\equiv\sqrt{r-r_h}$, the solution of the EOM for $R_{2,n}(u)$ is:
\begin{eqnarray}
\label{solution-high-T}
& & R_{2, n}(u) = u^{\Lambda}\left[c_1 U\left(\mu_1, \mu_2, \mu_3 u\right)+c_2 L_{-\mu_1}^{\mu_2-1}\left(\mu_3 u\right)
\right],
\end{eqnarray} 
where $\Lambda, \mu_{1, 2, 3}$ are defined in (\ref{arguments-U}), and $p = M_{\rm Mesino}\frac{r_h}{\sqrt{g_s N}}$ \footnote{ Glueball and meson masses at high temperatures were obtained respectively in \cite{IITR-glueballs} and \cite{V+A+K} in units of $\frac{r_h}{\sqrt{g_s N}}$.} is contained in the ${\cal O}\left(\frac{\beta}{N}\right)$ term in $\mu_3$, which hence remains undetermined as ${\cal O}\left(\frac{\beta}{N}\right)$ terms are dropped (see footnote {\bf 1}). One can show that  $\lim_{u\rightarrow0}u^{\Lambda }c_1 U\left(\mu_1, \mu_2, \mu_3 u\right)$ is singular. One hence can not impose Dirichlet or Neumann boundary condition at $r=r_h$ if $c_2=0$. Now,
{\footnotesize
\begin{eqnarray}
\label{LaguerreLu0}
& & L_{-\mu_1}^{\mu_2-1}(u) = \frac{\Gamma(\mu_2-\mu_1)}{\Gamma(1-\mu_1)\Gamma(\mu_2)} - \frac{\Gamma(\mu_2-\mu_1)}{\Gamma(-\mu_1)\Gamma(\mu_2+1)}u + {\cal O}(u^2),
\end{eqnarray} 
}
implying $\lim_{u\rightarrow 0}u^\Lambda c_2 L_{-\mu_1}^{\mu_2-1}(\mu_3 u)=0$, implying the Dirichlet boundary condition is identically satisfied $\forall M_{\rm Mesino}$ including very large $M_{\rm Mesino}$.  It is extremely non-trivial that the $\mu_i$s receive no ${\cal O}(\beta)$ corrections up to ${\cal O}\left(\frac{\beta}{N^{\alpha_{\mu_i}}}\right),\ \alpha_{\mu_i}\ge1$ - see (\ref{arguments-U}). 

The absence of ${\cal O}(R^4)$ corrections is essentially a reflection of the fact that the $SL(2, Z)$ completion of the effective $R^4$ interaction terms in type IIB supergravity leads to an interesting non-renormalization theorem that forbids perturbative corrections beyond one loop in the zero-instanton sector \cite{Green and Gutperle and Vanhove}.

What we now address in section \ref{Enhancement-mass-scale} is how an $N$-enhancement ($\equiv N$-hancement) of the mass scale $M_{\rm KK} = \frac{r_0}{\sqrt{4 \pi g_s N}}$ \cite{Vikas+Gopal+Aalok} is obtained which therefore explains how one could obtain supermassive $M_{\rm Mesino}$.

\section{Generation of $N$-hanced Mass Scale for $T<T_c$}
\label{Enhancement-mass-scale}
In this section, starting from the $D=11$ supergravity Einstein's field equations in the presence of four-form $G$ fluxes of ${\cal M}$-theory \footnote{One can show that ``$E_8$''-dependent terms in the same are subdominant as compared to the ``$J_0$''-dependent terms \cite{OR4}.}  - the first in (\ref{eoms}) (also given in \ref{Applications_MQGP}) - we explicitly show the generation of an $N$-enhanced ($\equiv ``N$-hanced'') mass scale, thereby providing the mechanism of generation of super-massive mesinos. 

Replacing the resolution parameter ``$a$'' of the blown-up $S^2$  by $a(r)$, substituting an ansatz: $a(r) = b + c^{\beta^0}(r-r_0) + \beta  {\cal A}^\beta(r)$ into ${\rm EOM}_{MN}$ in (\ref{eoms}) ($b$ being a ``bare'' resolution parameter) and estimating $r_0\sim e^{-\kappa_{r_0}N^{1/3}}$ \cite{Bulk-Viscosity-McGill-IIT-Roorkee}, near the $\psi=2n\pi, n=0, 1, 2$-coordinate patch, yields the following:
\begin{enumerate}

\item
\begin{eqnarray}
\label{EOMtt}
& & {\rm EOM}_{tt,x^1x^1/x^2x^2}:\ b\sim \kappa_{tt/x^ix^i/rr} N^{10/9}e^{-\kappa_{r_0}(N)\left(3 + 0.5 \kappa_{r_0}(N)N^{1/3}\right)N^{1/3}}r_0;\nonumber\\
& & {\cal A}^\beta(r) \sim e^{\frac{c^\beta}{b}r} {\cal C}_1, 
\end{eqnarray}
with $\kappa_{tt/x^ix^i/rr}\gg1,\ \kappa_{r_0}(N=10^2)=\frac{1}{{\cal O}(1)}-{\cal O}(1)$, one obtains $b\gg r_0$ and in principle an $r_0$-independent true bare resolution parameter proptional to $\beta$; ${\rm EOM}_{x^3x^3}$ near $r=r_0$ does not constrain $b$.

\item
\begin{eqnarray}
\label{EOMrr}
& & {\rm EOM}_{rr}:\ b\sim\tilde{\kappa}_{rr}N^{11/9}e^{\kappa_{r_0}N^{1/3}+1.25\sqrt{1.57 + 0.55\log r_0 - 0.5\left(\log r_0\right)^2}}r_0;
\end{eqnarray}
and for an appropriate $\kappa_{r_0}\sim\frac{1}{{\cal O}(1)}: 1.57 + 0.55\log r_0 - 0.5\left(\log r_0\right)^2>0$, and $N\sim10^2$, one regains the result for $b$ as obtained in the first equation in (\ref{EOMtt}).

\item
{\footnotesize
\begin{eqnarray}
\label{EOM66}
& & \hskip -0.8in {\rm EOM}_{\theta_1\theta_1}:\ a(r)\ {\rm determined\ by} :- \nonumber\\
& & \hskip -0.8in \frac{1323 \sqrt[5]{N} \alpha _{\theta _1}^2}{256 \alpha _{\theta _2}^2}-\frac{729 {g_s}^3 M^2
   \left(\frac{1}{N}\right)^{6/5} {N_f}^2 \left(2187 \alpha _{\theta _1}^6+270 \sqrt{6} \alpha _{\theta _2}^2 \alpha_{\theta _1}^3+50 \alpha _{\theta _2}^4\right) a(r)^3 \log ^3({r_0}) \left(2 r {r_0} \log ({r_0}) a'(r)+a(r)
   (r_0-r) \log ({r_0})\right)}{16 \pi ^3 r_0^5 \alpha _{\theta _1}^2 \alpha _{\theta _2}^2}=0,\nonumber\\
& &     
\end{eqnarray}   
}   
whose solution is given by:
{\footnotesize
\begin{eqnarray}
\label{a-EOM66-solution}
& & a(r) = \Biggl(\frac{864 c_1 {g_s}^3 M^2 {N_f}^2 \Sigma e^{\frac{2 r}{r_0}} \log ^4({r_0})-98 \pi ^3
    N^{7/5} r {r_0}^5 \alpha _{\theta _1}^4-49 \pi ^3
   N^{7/5} {r_0}^6 \alpha _{\theta _1}^4}{{g_s}^3 M^2 {N_f}^2 r^2 \Sigma \log^4({r_0})}\Biggr)^{1/4},\nonumber\\
& & \sim c_1\frac{e^{\frac{r}{2r_0}}}{\sqrt{r}} \sim \frac{c_1}{\sqrt{r_0}}\left[1 + {\cal O}\left(\frac{(r-r_0)^2}{r_0^2}\right)\right],   
\end{eqnarray}
}
where $\Sigma\equiv \left(2187 \alpha _{\theta _1}^6+270 \sqrt{6} \alpha _{\theta _2}^2 \alpha
   _{\theta _1}^3+50 \alpha _{\theta _2}^4\right)$. Recalling that $r_0\sim e^{-\kappa_{r_0}N^{1/3}}$, we reinterpret (\ref{EOM66}) as  $a(r\sim r_0)\sim c_1 e^{\frac{3\kappa_{r_0}N^{1/3}}{2}} r_0$, where for compatibity with (\ref{EOMtt}) and (\ref{EOMrr}), one may choose an $N$-dependent $c_1\sim N^{(10-11)/9}e^{-\gamma N^{1/3}}$ for an appropriate $\gamma$.

\item ${\rm EOM}_{\theta_1\theta_2}$
\begin{eqnarray}
\label{EOM67}
& & \lambda_3 a(r)^3 \left(a(r)-a'(r)\right)+\lambda_1 a(r)^4+\frac{\lambda_2 \left(36 a(r)^2 \log
   ({r_0})+{r_0}\right)}{{r_0}^2-3 a(r)^2}=0,
\end{eqnarray}
where,
\begin{eqnarray}
\label{defs-EOM67}
& & \lambda_1 \equiv -\frac{243 {g_s}^3 M^2 \left(\frac{1}{N}\right)^{11/10} N_f ^2 \left(2187 \alpha _{\theta
   _1}^6+270 \sqrt{6} \alpha _{\theta _2}^2 \alpha _{\theta _1}^3+50 \alpha _{\theta _2}^4\right) \log
   ^4({r_0})}{8 \pi ^3 {r_0}^4 \alpha _{\theta _1} \alpha _{\theta _2}^3},\nonumber\\
   & & \lambda_2 \equiv -\frac{1323 N^{3/10} {r_0} \alpha _{\theta _1}^3}{256 \alpha _{\theta _2}^3 (\log N -9 \log
   ({r_0}))},\nonumber\\
& & \lambda_3 \equiv -\frac{729 {g_s}^3 M^2 \left(\frac{1}{N}\right)^{11/10} N_f ^2 \left(2187 \alpha _{\theta
   _1}^6+270 \sqrt{6} \alpha _{\theta _2}^2 \alpha _{\theta _1}^3+50 \alpha _{\theta _2}^4\right) \log
   ^4({r_0})}{8 \pi ^3 {r_0}^3 \alpha _{\theta _1} \alpha _{\theta _2}^3}.   
\end{eqnarray}
Defining,
\begin{eqnarray}
\label{root}
& & \Lambda \equiv \frac{2^{5/6} \sqrt{{g_s}} \sqrt[3]{M} \sqrt[3]{N_f } {r_0}^2 \sqrt[6]{2187 \alpha _{\theta
   _1}^6+270 \sqrt{6} \alpha _{\theta _2}^2 \alpha _{\theta _1}^3+50 
   \alpha_{\theta _2}^4} \log^{\frac{25}{6}}({r_0})}{9 \sqrt[3]{7} \sqrt{3 \pi } N^{7/30} \alpha _{\theta _1}^{2/3}},
\end{eqnarray}
$a(r)$ is given by:
\begin{eqnarray}
\label{a-solution}
& & \sqrt{{\Lambda}+\exp \left(\frac{2 (r+\lambda_3 c_1) \left({\Lambda}
   (\lambda_1+\lambda_3) \left(2 {r_0}^2-9 {\Lambda}\right)+36 \lambda_2 \log
   ({r_0})\right)}{\lambda_3 {\Lambda} \left({r_0}^2-3 {\Lambda}\right)}\right)}\nonumber\\
& & \sim \sqrt{\Lambda} \sim \frac{\sqrt[4]{{g_s}} \sqrt[6]{M} \sqrt[6]{N_f }  \sqrt[12]{2187 \alpha _{\theta
   _1}^6+270 \sqrt{6} \alpha _{\theta _2}^2 \alpha _{\theta _1}^3+50 \alpha _{\theta _2}^4} \log^{\frac{25}{12}}({r_0})}{N^{7/60} \alpha _{\theta _1}^{1/3}}r_0.   
\end{eqnarray}

\item
${\rm EOM}_{\theta_1x}$:
\begin{eqnarray}
\label{EOM68}
& & b\sim N^{23/36} e^{\frac{1}{6}\kappa_{r_0}N^{1/3}\left(9\kappa_{r_0}N^{1/3} + \log N\right)}r_0.
\end{eqnarray}

\item
${\rm EOM}_{\theta_1y}$:
\begin{eqnarray}
\label{EOM69}
& & b \sim N e^{\frac{3}{2}\kappa_{r_0}^2N^{2/3}}r_0.
\end{eqnarray}

\item
${\rm EOM}_{\theta_2x}$:
\begin{eqnarray}
\label{EOM78}
& & b \sim \kappa_{\theta_2y}N^{10/9}e^{-\kappa_{r_0}^2N^{2/3}+4\kappa_{r_0}N^{1/3}}r_0,\ \kappa_{\theta_2x}\gg1.
\end{eqnarray}

\item
${\rm EOM}_{\theta_2y}$:
\begin{eqnarray}
\label{EOM79}
& & b\sim N^{10/9}e^{-3\kappa_{r_0}N^{1/3}+\kappa_{r_0}^2 N^{2/3}}r_0.
\end{eqnarray}

\item
${\rm EOM}_{\theta_2z}$:
\begin{eqnarray}
\label{EOM710}
& & b\sim N^{10/9}e^{\kappa_{r_0}^2N^{2/3} - 3\kappa_{r_0}N^{1/3}}r_0.
\end{eqnarray}

\item
${\rm EOM}_{xz/yy/yz/zz}$:
\begin{eqnarray}
\label{EOM89-810-99-910-1010}
& & b\sim N^{10/9} e^{\kappa_{r_0}^2N^{2/3}-6\kappa_{r_0}N^{1/3}}r_0.
\end{eqnarray}

\item
${\rm EOM}_{x^{10}x^{10}}$:
\begin{equation}
\label{EOM1111}
b\sim N^{10/9} r_0.
\end{equation}

\end{enumerate}

We therefore see that the ``bare resolution parameter'' $b$  given by: 
\begin{eqnarray}
\label{bare}
& & b \sim N^{1+\frac{1}{{\cal O}(1)}} r_0;\ \ \  a^\beta(r) = {\cal C} e^{\frac{c_{\rm linear}}{b}r},\ {\cal C}\equiv{\rm constant}.
\end{eqnarray}
One hence can not obtain an $r_0$-independent ``$b$''. {\it One thus sees an $N$-hancement of the effective KK mass scale $M_{KK}$ (from $M_{KK}$ to 
$M_{KK}^{\rm eff}\sim N^{1 + \frac{1}{{\cal O}(1)}}M_{KK}$)  arising from the construction of  SYZ type IIA mirror of the non-K\"{a}hler type IIB dual \cite{metrics} of thermal QCD-like theories, as well as the generation of a one-parameter (${\cal C}$) family of $r_0/M_{KK}$-independent bare resolution parameter at ${\cal O}(R^4)$ in the ${\cal M}$-theory uplift involving a $G_2$-structure wherein ${\cal C}$ can be made appropriately large.} These are the pair of reasons for generating super-massive mesinos in the fermionic sector in the string/${\cal M}$ theory duals of thermal QCD at finite $N$ in \cite{MQGP}, \cite{OR4}.

\section{Non-Interacting Mesinos}
\label{No-mesino-meson-interaction}

Given that we have seen in \ref{Fermion} that supermassive mesinos, unlike \cite{SS} (see \cite{mesino-SS}), {\it are} permissible in the type IIA holographic dual \cite{MQGP} at intermediate coupling \cite{OR4} of realistic thermal QCD-like theories, this already explains why mesinos have thus far not been observed near the EW scale. In this section, we will further show that mesino-mesino-single-($\rho/\pi$)meson interactions, unlike \cite{SS} (see \cite{mesino-SS}), vanish identically in the aforementioned type IIA holographic dual. Considering fluctuations of the vector mesons $A_{\mu, r}\rightarrow A_{\mu, r}^{(0)} + \delta A_{\mu, r}$ with $A_{\mu=t}^{(0)}$ being the only non-zero background value (see \ref{Fermion}) which can be shown to be tunable so that $|F_{rt}^{(0)}|\ll1$, implying one need only consider terms linear in $F_{\rm IIA}^{(0)}$ \footnote{Use is made of $i^*B_{\alpha_1\alpha_2} = \delta^{[\theta_1}_{\alpha_1}\delta^{\theta_2]}_{\alpha_2}B_{\theta_1\theta_2}$ and consequently, ${\cal F}_{rt}=F_{rt}$.} which are contained (recalling from section \ref{Fermion}, {\footnotesize $\sqrt{-{\rm det}(i^*g^{\rm IIA}+\mathcal{F}_{\rm IIA}^{(0)})} \ll1,\ {\cal F}_{IIA} = i^*B_{IIA} + F$} in the large-$N$ limit) in :
{\footnotesize
\begin{eqnarray}
\label{DBI-F-ii}
S_{D_6}^{f}=\frac{T_{D_6}}{2} \int d^7\xi e^{-\Phi^{\rm IIA}}\  \overline{\Theta}\left({\frac{\Gamma^{\beta_1.....\beta_{7}}\mathcal{F}_{\rm IIA}^{(0)}\ _{\beta_{6} \beta_{7}}\Gamma_{\beta_1.....\beta_{5}}^{\ \ \ \ \ \ \ \ \gamma} D_\gamma\Gamma_{(10)}}{\sqrt{-{\rm det}(i^*g^{\rm IIA}+\mathcal{F}_{\rm IIA}^{(0)})} }} \right)\Theta.
\end{eqnarray}
}
Considering fluctuations in the background gauge field in (\ref{DBI-F-ii}) and retaining terms linear in the same yields:
{\footnotesize
\begin{eqnarray}
\label{DBI-F-iii}
\delta S_{D_6}^{f} \sim T_{D_6} \int_{\Sigma_{(7)}} d^4x dr d\theta_2 d\tilde{y} e^{-\Phi^{\rm IIA}} \overline{\Theta}\left(\frac{4\Gamma^{\beta_1.....\beta_{7}}\delta \mathcal{F}^{\rm IIA}_{\beta_{6} \beta_{7}}\Gamma_{\beta_1.....\beta_{5}}^{\ \ \ \ \ \ \ \ \gamma} D_\gamma\Gamma_{(10)}}{\sqrt{-{\rm det}(i^*g^{\rm IIA}+\mathcal{F}_{\rm IIA}^{(0)})}} \right)\Theta.
\end{eqnarray}
}
The next step is to perform the KK expansion of $\delta\mathcal{F}^{\rm IIA}_{\alpha \beta}$ and decompose spinors along $M_4$ and internal directions, and by integrating over the $\theta_2$ and $\tilde{y}$ we will get mesino-mesino-meson interaction action with couplings given in terms of radial integrals of the radial profile functions of the mesino and mesons. The usual KK expansion ansatz \cite{Vikas+Gopal+Aalok} is: 
\begin{equation}
\label{KK-i}
\delta A_\mu(x^\mu,r)=\sum_{n=1}^{\infty} \rho_\mu^{(n)}(x) \psi_n(r),
\end{equation}
 and 
\begin{equation}
\label{KK-ii} 
 \delta A_r(x^\mu,r)=\sum_{n=0}^{\infty} \pi^{(n)}(x) \phi_n(r),
\end{equation} 
  implies 
\begin{equation}
\label{KK-iii}  
  \delta F_{\mu \nu}=\sum_{n=1}^{\infty} \tilde{F}_{\mu \nu}^{(n)}(x) \psi_n(r),
\end{equation}  
   and 
\begin{equation}
\label{KK-iv}   
   \delta F_{\mu r}=\sum_{n=0}^{\infty} \partial_\mu \pi^{(n)}(x^\mu) \phi_n(r)- \sum_{n=1}^{\infty}\rho_\mu^{(n)}(x) \dot{\psi}_n(r).
\end{equation}   
We will keep the $n=1$ term for the vector fluctuation and $n=0$ for the $A_r(x^\mu,r)$; hence, the degrees of freedom are $\rho$ vector meson and $\pi$ meson.   Using the KK decomposition of $\delta F_{\mu \nu}$ and $\delta F_{\mu r}$,  (\ref{DBI-F-iii}) simplified as follows:
{
\begin{eqnarray}
\label{DBI-F-v}
& & 
S_{D_6}^{int} \sim T_{D_6} \int_{\Sigma_{(7)}}\Biggl[ \frac{e^{-\Phi^{\rm IIA}}}{\sqrt{-{\rm det}(i^*g^{\rm IIA}+\mathcal{F}_{\rm IIA}^{(0)})}} \overline{\Theta}\Biggl(\Gamma^{\beta_1.....\beta_{5}\mu \nu}\delta{\tilde{F}}_{\mu \nu} \psi(r)\Gamma_{\beta_1.....\beta_{5}}^{\ \ \ \ \ \ \ \ \gamma} D_\gamma  
\nonumber\\
& & \hskip 0.4in +\Gamma^{\beta_1.....\beta_{5}\mu r}\left(\partial_\mu \pi(x^\mu) \phi(r)- \rho_\mu(x^\mu) \dot{\psi}(r)\right)\Gamma_{\beta_1.....\beta_{5}}^{\ \ \ \ \ \ \ \ \gamma} D_\gamma \Biggr)\Theta \Biggr].
\end{eqnarray}
} 
\noindent 
Using the decomposition of the ten-dimensional gamma matrices \cite{Kirsch}:
\begin{eqnarray}
\label{gamma}
& & \Gamma^{\underline{A}=\underline{t},\underline{x^{1, 2, 3}}, \underline{r}} =\sigma_y\otimes{\bf 1}_4\otimes\gamma^{\underline{A}},\ \ \  \Gamma^{\underline{a}=5,...,9} = \sigma_x\otimes\gamma^{\underline{a}}\otimes{\bf 1}_4, 
\end{eqnarray} 
with 
\begin{eqnarray}
& & 
\left\{\gamma^{\underline{A}},\gamma^{\underline{B}}\right\} = -2 \eta^{\underline{AB}}, \nonumber\\
& &\left\{\gamma^{\underline{a}},\gamma^{\underline{b}}\right\} = -2\delta^{\underline{ab}}.
\end{eqnarray}
The ten-dimensional chirality matrix is defined as:
\begin{eqnarray}
& & \Gamma^{(10)} = \sigma_z\otimes{\bf 1}_4\otimes{\bf 1}_4 .
\end{eqnarray}
The positive-chirality ten-dimensional $\Theta$ can hence be decomposed into
\begin{eqnarray}
\Theta = \uparrow\otimes\chi_{M_5(x^{0, 1, 2, 3}, r)}\otimes\psi_{\tilde{M}_5(\theta_{1,2}, \phi_{1,2}, \psi)},
\end{eqnarray}
 where $\psi_{\tilde{M}_5(\theta_{1,2}, \phi_{1,2}, \psi)}$ further splits into $\psi_{\tilde{M}_5} = \psi_{S^2_{\rm squashed}}\otimes \psi_{S^3_{\rm squashed}}$. Looking at the second fermionic bilinear in (\ref{DBI-F-v}):
\begin{eqnarray} 
&  &
  {\bar\Theta}\Gamma^{\beta_1...\beta_5 t r}\Gamma_{\beta_1...\beta_5}^{\ \ \ \ \  \ \gamma}D_\gamma\Theta (\beta_{i=1,...,5}=x^{0, 1, 2, 3}, \theta_2, \tilde{y};\ \gamma = t, r) \sim {\bar\Theta} \Gamma^{t} D_r\Theta + {\bar\Theta} \Gamma^{r} D_t\Theta.
\end{eqnarray}  
   Now, the non-vanishing ${\bar\Theta}\Gamma^{\underline{X}_1....\underline{X}_p}\Theta$ involving Majorana-Weyl spinor $\Theta$ requires $p = 3, 7$ \cite{mesino-SS}. One can further show that the most dominant spin-connection component of the type $\omega_r^{\underline{a b}}$ is $\omega_r^{\underline{7}\underline{9}}$ and only non-vanishing spin-connection component of the type $\omega_t^{\underline{a}\underline{b}}$ is $\omega_t^{\underline{x^0}\underline{r}}$. Therefore, using (\ref{gamma}):
\begin{eqnarray}
& & {\bar\Theta}\Gamma^{\beta_1...\beta_5 t r}\Gamma_{\beta_1...\beta_5}^{\ \ \ \ \ \  \gamma}D_\gamma\Theta\sim{\bar\Theta}\Gamma^{\underline{t}}\omega_r^{\underline{7 9}}\Gamma^{\underline{7 9}}\Theta \propto \langle \uparrow | \sigma_y |\uparrow\rangle=0.
\end{eqnarray}   
    Also, 
\begin{eqnarray}    
    & & \overline{\Theta}\Gamma^{\beta_1.....\beta_{5}\mu \nu}\delta{\tilde{F}}_{\mu \nu} \psi(r)\Gamma_{\beta_1.....\beta_{5}}^{\ \ \ \ \ \ \ \ \gamma} D_\gamma\Theta = 0, 
\end{eqnarray}    
    as $\mu, \nu \in x^{1, 2, 3}$ and thus using (\ref{gamma}):
    \begin{eqnarray}
   & & \hskip -0.3in \overline{\Theta}\Gamma^{k x^0 r \theta_2 \tilde{y}i j}\delta{\tilde{F}}_{i j} \psi(r)\Gamma_{k x^0 r \theta_2 \tilde{y}}^{\ \ \ \ \ \ \ \ i} D_i\Theta (i\neq j\neq k = x^{1, 2, 3}) \sim \delta{\tilde{F}}_{i j}\overline{\Theta}\Gamma^{i j}\Gamma^l \partial_l\Theta (l\neq k)\ \propto \langle \uparrow | \sigma_y |\uparrow\rangle=0.
\end{eqnarray}   
Hence, no mesino-mesino-$\rho$/$\pi$-meson vertex is generated. Together with what was argued earlier that one could have a supermassive mesino, this suggests the ``WISP''(Weakly Interacting Supermassive Particle)y nature of the non-supersymmetric mesino, and consequently resolves the tension between actual QCD and top-down holographic QCD \cite{mesino-SS}. 
\section{Top-Down $m_{\rm quark}\langle {\bar q}q \rangle$ Non-Renormalization up to ${\cal O}(R^4)$}
\label{mq_chiral_cond_RG-inv_OR4-ind}

The ${\cal O}(R^4)$ corrections to the ${\cal M}$-theory dual's metric are vanishing small in the UV \cite{Gopal+Vikas+Aalok}. The EOM of the flavor $D6$-branes' embedding, $\tilde{z} = \tilde{z}(r)$ in the IR arising from the DBI action for the flavor $D6$-branes with world volume $\Sigma_7\left(S^1_t\times\mathbb{R}^3\times \mathbb{R}_{>0}\times S^2_{\rm squashed}\right)$ embedded via $i:\Sigma_7\hookrightarrow S^1_t\times \mathbb{R}^3\times_w M_6$ [$w\equiv$ warped product] effected by $\tilde{z} = \tilde{z}(r)$ in a non-K\"{a}hler warped squashed resolved conifold $M_6$ in the type IIA mirror of the UV-complete type IIB dual \cite{metrics} of thermal QCD-like theories, using the induced metric on flavor $D6$-branes of (\ref{induced-metric-D6}), NS-NS $B^{\rm IIA}$ of (\ref{B-IIA-diag-diag}), can be shown to yield: $\tilde{z}=$constant, inclusive of ${\cal O}(\beta)$ corrections. 

The DBI action in the UV is given by (disregarding overall $r$-independent factors, and hence the $\sim$):
\begin{equation}
\label{L-DBI}
 {\cal L}_{\rm DBI}^{\rm D6} \sim  \frac{ r^2 \sqrt{\frac{4 \pi  \sqrt{{g_s}} r^2 \alpha _{\theta _2}^3 \left(6 a^2+r^2\right)}{9 a^2+r^2}+3
   \sqrt{3 \pi } N^{2/5} \left(r^4-{r_h}^4\right) \tilde{z}'(r)^2}}{{g_s}^{3/4}},
\end{equation}
and consequently the $\tilde{z}(r)$ EOM: $\frac{\delta {\cal L}_{\rm DBI}^{\rm D6}}{\delta \tilde{z}'(r)} = {\cal K}$ (constant) in the UV yields:
\begin{equation}
\label{deltaz'}
\tilde{z}'(r) = \frac{{\cal C}}{r^5},
\end{equation}
${\cal C}$ being a constant (subsuming $g_s$- and $N$-dependent factors). One hence obtains\footnote{$\tilde{\cal C}_2=\frac{{\cal C}}{4}$.}: 
\begin{equation}
\label{deltaz}
\tilde{z}(r) = {\cal C}_1 -\frac{\tilde{\cal C}_2}{r^4}\stackrel{r\in{\rm UV}}{\longrightarrow}{\cal C}_1.
\end{equation}
As $\tilde{z}(r)$ is dimensionless, ${\cal C}_1$ will hence also be so, and ${\cal C}_2$ will have a mass dimension of four (in units of ${\cal R}_{D5/\overline{D5}}=D5-\overline{D5}$-separation). By looking at fluctuations: $\tilde{z}\rightarrow\tilde{z}+\delta \tilde{z}$ in the DBI action (no mass term $\left(\delta\tilde{z}\right)^2$ is generated) one can show that in the UV and in the $\psi=2n\pi, n=0, 1, 2$-coordinate patches and by working near, e.g., $\left(\theta_1, \theta_2\right)=\left(\frac{\alpha_{\theta_1}}{N^{1/5}}, \frac{\alpha_{\theta_2}}{N^{3/10}}\right)$ \cite{NPB}, \cite{Bulk-Viscosity-McGill-IIT-Roorkee} (consistent with the $\mu_{\rm Ouyang}|\ll1$-limit of the flavor $D7$-branes in the parent type IIB dual \cite{metrics}):
\begin{equation}
 \delta\tilde{z} \stackrel{\rm UV}{\longrightarrow}  c_1 + \frac{{\cal C}_2}{r^4} + {\cal O}\left(\left(\frac{{\cal C}_2}{r}\right)^{12}\right).
 \end{equation}
Again, we see that the mass dimension of the coefficient ${\cal C}_2$ of $\frac{1}{r^4}$ is four (and $c_1$ is dimensionless). Given that one obtains an $AdS_5$ in the UV, the coefficient of $\frac{1}{r^4}$ for the massless fluctuation $\delta\tilde{z}$ is identified with a chiral condensate \cite{Witten-chiral-cond_mq}, we conjecture that ${\cal C}_2$ is the top-down holographic analog of the mass-dimension-four $m_q\langle {\bar q}q\rangle$. As the ${\cal O}(R^4)$ corrections are vanishingly small in the UV \cite{Vikas+Gopal+Aalok}, $c_1$ and ${\cal C}_2$ receive no ${\cal O}(R^4)$ corrections. This is the top-down holographic analog of the RG-invariance of $m_q\langle {\bar q}q\rangle$  \cite{mqqbarq-RGI}.

\section{Universality in Particle Wave Functions in the IR}
\label{U+L_universality}

An intriguing universality in the wave functions of the following particle spectroscopies is noticed.

\noindent {\bf Glueballs} \cite{IITR-glueballs}: 
\begin{itemize}
\item
$0^{-+}$ glueball: The EOM of the type IIA RR $A$ fluctuation (to which tr$F\wedge\tilde{F}$ couples via the type IIA $D4$-brane with world volume $\Sigma_{1,4}$) term WZ term $\int_{\Sigma_{1,4}} A\wedge {\rm tr} F\wedge\tilde{F}$)  $\partial_\nu\left(\sqrt{g^{\rm IIA}}g_{\rm IIA}^{\mu\sigma}g_{\rm IIA}^{\nu\rho}\left(\partial_{[\sigma}A_{\rho]}\right)\right)=0$ (where $\mu,\nu,...=a(\equiv 0,1,2,3),r,\alpha(\equiv 5,...,9)$ and it was assumed $A_\mu=\delta^{\theta_2}_\mu a_{\theta_2}(r)e^{i k\cdot x}, k^2=-m^2$ as the fluctuation about the type IIA $A_1$ that was worked out in \cite{MQGP}).

\item
$0^{--}$ glueball: The EOM for the  fluctuation in the type IIB $A_{MN} = B_{MN} + i C_{MN}$ (that figures in the Weiss-Zumino term $A^{\mu\nu}d^{abc}{\rm Tr}\left(F_{\mu\rho}^aF^{b\ \rho}_\lambda F^{c\ \lambda}_\nu\right)$), $\delta A^{MN} = \delta^M_2\delta^N_2 \delta A_{23}$, \\
$\partial_\mu\left(\sqrt{-g}g^{22}g^{33}g^{\mu\nu}\partial\delta A_{23}\right)=0$.

\item
$1^{++}$ glueball: The EOM for the radial profile function of the vector-type ${\cal M}$-theory metric perturbation $h_{ti} = h_{it} = g_{x^1x^1}G(r) e^{i k x^1}, i=x^2, x^3: R^{(1)}_{\mu\nu}\approx 0,\ R^{(1)}_{\mu\nu}$ denoting the first-order fluctuations in the Ricci tensor as a consequence of linear metric perturbations.

\end{itemize}

\noindent {\bf Mesons} \cite{V+A+K}: Working with the redefined radial variable $Z: r = r_h e^Z$,  after integrating out the blown-up $S^2_{\rm squashed}$ in the DBI action of the flavor type IIA $D6$-branes and  KK reduction of the gauge field {\footnotesize $A_\mu(x^\mu,Z)=\sum_{n=1}B_\mu^{(n)}(x^\mu)\alpha^{\{\mu\}}_n(Z), \mu = t, x^{i=1, 2, 3}$}, the terms in the DBI action quadratic in the gauge field fluctuations are: {\footnotesize $\int d^4x dZ \left({\cal V}_2(Z)F^{(n)}_{\mu\nu}F^{\mu\nu}_{(n)}\alpha_m^{\left\{\mu\right\}}(Z)\alpha_n^{\left\{\mu\right\}}(Z) + {\cal V}_1(Z)B^{(m)}_\mu B^{(n)}_\nu\dot{\alpha}_m^{\left\{\mu\right\}}\dot{\alpha}_n^{\left\{\mu\right\}}\right)$}. The EOM for the radial profile  {\footnotesize $\alpha^{\{i\}}_m(Z)$} is : {\footnotesize $\frac{d}{dZ}\left({\cal V}_1(Z)\dot{\alpha}_m^{\{i\}}\right) + 2{\cal V}_2(Z){\cal M}_{(m)}^2\alpha_m^{\{i\}} = 0$}, where \\ {\footnotesize
$ {\cal V}_1(Z) = e^{-\Phi^{IIA}}\sqrt{h}g^{ZZ}\sqrt{{\rm det}_{\theta_2,\tilde{y}}\left(i^*(g+B)\right)}\sqrt{{\rm det}_{\mathbb{R}^{1,3},Z}(i^*g)}$} and {\footnotesize $ {\cal V}_2(Z) =  e^{-\Phi^{IIA}}\frac{h}{2}\sqrt{{\rm det}_{\theta_2,\tilde{y}}\left(i^*(g+B)\right)}\sqrt{{\rm det}_{\mathbb{R}^{1,3},|Z|}(i^*g)}$}. The solution of $\alpha^{\{i\}}$ is given in terms of the Tricomi Hypergeometric and associated Laguerre functions.

\noindent {\bf Graviton} \cite{Gopal+Aalok}: In the context of obtaining the Page curve of an eternal black hole from the ${\cal M}$-theory dual containing a black-hole in the ETW(End of The World)-``brane'' (a hypersurface $AdS_4^\infty\times_w M_6, \times_w$ implying a warped product, with $G_4$ fluxes threading a homologous sum of four-cycles $S^3_{\rm squashed}\times[0,1]$ and $S^2_{\rm squashed}\times S^2_{\rm squashed}$ in $M_6=M_5(\theta_{1,2}, \phi_{1,2},\psi)\times S^1(x^{10})\hookrightarrow M^{SU(4)/Spin(7)}_8(t,r,\theta_{1,2}, \phi_{1,2},\psi,x^{10})$, with a finite ``tension'') coupled to a non-conformal QCD bath in the doubly holographic approach, the massless graviton wavefunction with the graviton localized on the ETW-brane trapped in a ``volcano''-like potential, is given in terms of the Tricomi Hypergeometric and associated Laguerre functions.

Solutions to the EOMs for the aforementioned field fluctuations/radial profile function are given in terms of the Tricomi Hypergeometric and associated Laguerre functions. The reason is that the relevant near-$r_h$ EOMs for $0^{-+}, 0^{--}, 1^{++}$-glueballs \cite{IITR-glueballs},  and the radial profile function of the graviton wave function \cite{Gopal+Aalok} are {\it all} of the type:
\begin{equation}
\label{universality-i}
(r-r_h)\xi''(r) + \left(b + c (r-r_h) \right)\xi'(r) + (f + (r-r_h)G)\xi(r) = 0,
\end{equation}
whose solution is given as:
{\footnotesize
\begin{eqnarray}
\label{universality-ii}
& & \hskip -0.4in  \xi(r\sim r_h) = e^{-\frac{1}{2} r \left(\sqrt{c^2-4 G}+c\right)}
   \Biggl[c_1 U\left(\frac{b \left(c+\sqrt{c^2-4
   G}\right)-2 f}{2 \sqrt{c^2-4 G}},b,\sqrt{c^2-4 G}
   (r-r_h)\right)  +c_2 L_{\frac{2 f-b
   \left(c+\sqrt{c^2-4 G}\right)}{2 \sqrt{c^2-4
   G}}}^{b-1}\left(\sqrt{c^2-4 G}
   (r-r_h)\right)\Biggr].\nonumber\\
& & 
\end{eqnarray}
}
In the context of the radial profile functions of vector mesons \cite{V+A+K}, and mesinos at $T>T_c$ in equation (\ref{ang-T_greater_than_Tc}), after appropriate coordinate redefinitions, the near-horizon (IR) solutions are also given in terms of the Tricomi Hypergeometric and Associate Laguerre functions.

\section{Summary}
\label{summary}
The immensely popular holographic QCD dual of \cite{SS} suffered from the longstanding problem that the Mesinos were nearly isospectral with the mesons, with non-vanishing/un(large-$N$-)suppressed mesino-mesino-meson interaction \cite{mesino-SS}- both in direct conflict with real QCD. What we show is that using the type IIA Strominger-Yau-Zaslow mirror of the UV-complete \cite{metrics} (unlike \cite{SS} which caters only to the IR) as constructed in \cite{MQGP} inclusive of ${\cal O}(R^4)$ corrections worked out in \cite{OR4}, not only is it possible to have super-massive mesinos that do not interact with the mesons, the results obtained (mesino wave function, mass, mesino-mesino-meson interaction) receive no ${\cal O}(R^4)$ corrections up to ${\cal O}\left(\frac{l_p^6}{N^\alpha}\right), \alpha\geq1$. Thus, the  {\it ``WISP''(Weakly Interacting Super-massive Particles)y mesinos and non-renormalization of their wave functions and mass up to ${\cal O}(R^4)$, together, apart from solving a longstanding problem, also provide a major and new insight into the fermionic sector of top-down holographic duals close to real thermal QCD.}

Further, the product of the quark mass and chiral condensate may be conjectured to correspond to the coefficient of the leading non-constant term in the flavor $D6$-branes' embedding's fluctuation, with the RG-invariance of the former \cite{mqqbarq-RGI} corresponding to the non-renormalization up to ${\cal O}(R^4)$ of the latter. In the end, we would also point out that there is a rather intriguing wave-function universality in the form of the appearance of (appropriate) Tricomi Hypergeometric and Associate Laguerre function in the glueball/meson/graviton (apart from mesinoic) spectroscopies.

\section*{Acknowledgements}

\vskip -0.2in

AM is partly supported by a Core Research Grant number SER-1829-PHY from the Science and Engineering Research Board, Govt. of India. GY is supported by a Senior Research Fellowship (SRF) from the Council of Scientific and Industrial Research, Govt. of India. We thank Nick Evans for a useful clarification. GY thanks the Infosys Foundation for the partial support at CMI.


\appendix



\section{Finite Quark Chemical Potential}
\setcounter{equation}{0}
\seceqaa
\label{finite-mu-baryon}

We explicitly show the generation of a finite quark chemical potential. From equation (\ref{ang-T_greater_than_Tc}) ($f(r)$ being valid $\forall r$), up to LO in N, $k^{\rm UV}(r)=1-3\frac{a^2}{r^2}, f(r) = \frac{2N^{2/5}r^6}{729\pi g_s \alpha_{\theta_1}^4\alpha_{\theta_2}^2}$ and integrating $\frac{\kappa\sqrt{k^{\rm UV}(r)}}{\sqrt{\kappa^2+f^2(r)}}$, one obtains:
 {\footnotesize
\begin{eqnarray}
\label{At-UV}
& & \hskip -0.5in A_t(r\in{\rm UV}) \sim \frac{1}{\sqrt{i \left(r^2-3 a^2\right)}}\nonumber\\
& & \hskip -0.5in \times\Biggl\{(-1)^{2/3} \sqrt[4]{3} a r \sqrt{1-\frac{3 a^2}{r^2}} \Biggl(F\left[\sin ^{-1}\left(\frac{3^{3/4} \sqrt[3]{\kappa }
   \sqrt[6]{\frac{\pi }{2}} \sqrt{i \left(r^2-3 a^2\right)} \sqrt[6]{g_s} \alpha _{\theta _1}^{2/3} \sqrt[3]{\alpha
   _{\theta _2}}}{{ar} \sqrt[15]{N}}\right)|\frac{1}{2} \left(1-i \sqrt{3}\right)\right]\nonumber\\
   & & \hskip -0.5in -\Pi \left[\frac{\frac{3 i
   2^{2/3} \kappa ^{2/3} \sqrt[3]{\pi } \sqrt[3]{g_s} \alpha _{\theta _1}^{4/3} \alpha _{\theta _2}^{2/3}}{a^2
   N^{2/15}}-2 (-1)^{5/6}}{2 \sqrt{3}};\sin ^{-1}\left(\frac{3^{3/4} \sqrt[3]{\kappa } \sqrt[6]{\frac{\pi }{2}} \sqrt{i
   \left(r^2-3 a^2\right)} \sqrt[6]{g_s} \alpha _{\theta _1}^{2/3} \sqrt[3]{\alpha _{\theta _2}}}{{ar}
   \sqrt[15]{N}}\right)|\frac{1}{2} \left(1-i \sqrt{3}\right)\right]\Biggr)\Biggr\}
   \nonumber\\
 & & \hskip -0.5in \sim \frac{3 \kappa \sqrt{\pi} \alpha _{\theta _1}^2 \alpha _{\theta _2}\left(1 -\frac{3   a^2}{r^2}\right)^{3/2} 
   \sqrt{g_s}}{\sqrt{2} a^2 \sqrt[5]{N}},  
\end{eqnarray} 
}
$F(\phi|\mu)\equiv\int_0^\phi\frac{d\alpha}{\sqrt{1 - m^2\sin^2\alpha}},$ being the incomplete elliptic integral of the first kind,  and $\Pi(\nu; \phi|\mu) \equiv \int_0^\phi\frac{d\alpha}{(1 - \nu^2\sin^2\alpha)\sqrt{1 - \mu^2\sin^2\alpha}}$ being the incomplete integral of the fourth kind, generating a finite quark chemical potential:
\begin{equation}
\label{chem-potential}
\mu = \frac{3 \sqrt{\pi } \kappa  \alpha _{\theta _1}^2 \alpha _{\theta _2} \sqrt{g_s}}{\sqrt{2} a^2 \sqrt[5]{N}}.
\end{equation}

\section{EOM-Related for Massive Mesinos}
\label{EOM-massive-mesinos}
\setcounter{equation}{0}
\seceqbb

The EOM for the radial profile $R_{2, n}(r)$ of the Mesino $\Theta$, as defined in equation (\ref{mode-ansatz-i}), is given by:
{\footnotesize
\begin{eqnarray}
\label{EOM-massive-i}
& &\frac{\Gamma^{\underline{1} \underline{5}} {E^r_{\underline{5}}}(r) R_{2, n}''(r)}{{E^t_{\underline{1}}}(r)}\Theta_{2,0} + R_{2, n}'(r) \left(\Gamma^{\underline{1} \underline{5}} \left(\frac{{E^r_{\underline{5}}}(r)}{{E^t_{\underline{1}}}(r)}\right)'-\frac{\Gamma^{\underline{1} \underline{2}}
   i {p}\  {E^{x^1}_{\underline{2}}}(r)}{{E^t_{\underline{1}}}(r)}-\frac{\Gamma^{\underline{1} \underline{5}} {\cal J}'(r)}{{\omega_r^{\underline{8}\ \underline{10}}}(r)}+2 \pi  i (2 n+1)
   T\right)\Theta_{2,0}\nonumber\\
 & & + R_{2, n}(r) \Biggl(-\frac{\pi ^2 \Gamma^{\underline{1} \underline{5}} (2 n+1)^2 T^2 {E^t_{\underline{1}}}(r)}{{E^r_{\underline{5}}}(r)}+\frac{\pi 
   \Gamma^{\underline{2} \underline{5}} (2 n+1) {p}\  T {E^{x^1}_{\underline{2}}}(r)}{{E^r_{\underline{5}}}(r)}+\frac{\Gamma^{\underline{1} \underline{2}} i {p}\  {E^{x^1}_{\underline{2}}}(r)
   {\cal J}'(r)}{{E^t_{\underline{1}}}(r) {\cal J}(r)}\nonumber\\
   & & +\frac{\Gamma^{\underline{1} \underline{5}} {p}\ ^2 {E^{x^1}_{\underline{2}}}(r)^2}{{E^t_{\underline{1}}}(r)^2}\omega_r^{\underline{a}\ \underline{10}}-\frac{\pi 
   \Gamma^{\underline{2} \underline{5}} (2 n+1) i {p}\  T {E^{x^1}_{\underline{2}}}(r)}{{E^t_{\underline{1}}}(r)}\frac{\omega_r^{\underline{a}\ \underline{10}}}{{\cal J}(r)}-\frac{\pi  i (2 n+1) T {\cal J}'(r)}{{\cal J}(r)}+{\cal J}(r)
   {\omega_r^{\underline{a}\ \underline{10}}}(r)\Gamma^{\underline{1}\underline{5}} - i \pi\left(\frac{E^{x^1}_{\underline{2}}}{E^t_{\underline{1}}}\right)'\Gamma^{\underline{1}\underline{2}}\Biggr)\Theta_{2,0}=0,\nonumber\\  
\end{eqnarray}
}
with $a=7, 8$ respectively for the TH, BH backgrounds with suitable aforementioned definitions for ${\cal J}(r)$. 

\begin{itemize} 
\item {\bf $T<T_c$}: Writing $M_{\rm Mesino} = \tilde{M}_{\rm Mesino}\frac{r_0}{\sqrt{g_s N}}$, the constants appearing in the Schr\"{o}dinger-like EOM (\ref{EOM-thermal-iii})-(\ref{EOM-thermal-v}), are given as under:
{\footnotesize
\begin{eqnarray}
\label{Cal-A-Bs}
& & \hskip -0.8in a_1 \equiv \frac{r_0^2}{\sqrt{3\pi g_s N}};\ a_2 \equiv \frac{23 r_0}{12\sqrt{3\pi g_s N}};\ b_1\equiv \frac{23 r_0}{12\sqrt{3\pi g_s N}};\ b_6 \equiv -\frac{23\sqrt{\pi g_s N}}{4\sqrt{3}r_0^3},\nonumber\\
& & \hskip -0.8in {\cal A}_{\Theta_2'} \equiv  -\frac{1511.7 \sqrt{{r_0}} \alpha _{\theta _2}}{{g_s}^{7/2} \kappa_2 \log N  M N^{2/5} {N_f}^2 \alpha _{\theta _1}^2},\nonumber\\
& &  \hskip -0.8in {\cal B}_{\Theta_2'} \equiv a_2 + \frac{2^{3/2}\tilde{M}_{\rm Mesino}\pi^{1/4}}{\left(g_s N\right)^{1/4}} + 2 i (2n+1)\pi T ,\nonumber\\
& &  \hskip -0.8in {\cal A}_{\Theta_2} \equiv \frac{-\frac{39.4 \tilde{M}_{\rm Mesino}}{\sqrt[4]{{g_s}} \sqrt[4]{N}}+T \left(n (-39.5 {a_6^{\beta}} {r_0} T-(125 i))-(62.5 i)-39.5 {a_6^{\beta}} n^2 {r_0} T-9.9 {a_6^{\beta}} {r_0} T\right)}{{r_0}},\nonumber\\
& &  \hskip -0.8in {\cal B}_{\Theta_2} \equiv \frac{M \left(\lambda_5^2 {g_s}^{15/4} \kappa_2^2 \log N ^2 \tilde{M}_{\rm Mesino} (0.1 n+0.1) N^{3/20} {N_f}^2
   {r_0}^4 T \alpha _{\theta _2}^2 \alpha _{\theta _1}^8-1.1 \lambda_5 {g_s}^{3/2} \kappa_2 \log N  \tilde{M}_{\rm Mesino}^2
   {N_f} {r_0}^2 \alpha _{\theta _2} \alpha _{\theta _1}^4+6.84 N^{3/5}\right)}{\lambda_5^2 \sqrt{{g_s}} \kappa_2 \log N 
   {r_0}^{11/2} \alpha _{\theta _1}^6 \alpha _{\theta _2}^3},
\nonumber\\
& & \hskip -0.8in {\cal C}_{\Theta_2} \equiv \frac{\frac{786.1 \tilde{M}_{\rm Mesino}}{\sqrt[4]{{g_s}} \sqrt[4]{N}}+T \left(n \left(-39.5 {b_6} {r_0}^2 T+(0.\, +2485.6
   i)\right)+(1242.8 i)-39.5 {b_6} n^2 {r_0}^2 T-9.9 {b_6} {r_0}^2 T\right)}{{r_0}^2},         
\end{eqnarray}
}
\noindent $\lambda_5$ being the parameter in terms of which the co-frames of the relevant non-K\"{a}hler six-folds were worked out in \cite{OR4}, $g_{\theta_2\theta_2}^{\rm IIA}(r\sim r_0)\sim\kappa_2\sqrt{g_s N}$ and,
\begin{eqnarray}
\label{a6inclbetadef}
& & \hskip -0.8in {a_6^{\beta}} \equiv \frac{\sqrt{3 \pi } \sqrt{{g_s}} \sqrt{N}}{{r_0}^2}+\frac{\beta  \sqrt{{g_s}} M \left(19683 \sqrt{6} \alpha _{\theta _1}^6+6642 \alpha _{\theta _2}^2 \alpha _{\theta _1}^3-40 \sqrt{6} \alpha
   _{\theta _2}^4\right) \log ^3({r_0})}{4374 \sqrt{\pi } \epsilon ^5 \log N ^4 N^{3/4} {N_f} {r_0}^4 \alpha _{\theta
   _2}^3}.
\end{eqnarray}

\item {\bf $T>T_c$}: Based on
{\footnotesize
\begin{eqnarray}
\label{omegar810}
& &\hskip -0.4in E^r_{\underline{5}} = \frac{\sqrt{\frac{9 a^2+r^2}{6 a^2+r^2}} \sqrt{r^4-{r_h}^4} \left(1-\frac{1}{2} \beta  \left({\cal C}_{zz}-2
   {\cal C}_{\theta_1z}+2 {\cal C}_{\theta_1x}\right)\right)}{\sqrt{2} \sqrt[4]{\pi } \sqrt{{g_s}} \sqrt{N} r},\nonumber\\
& & \hskip -0.4in E^t_{\underline{1}} = \frac{\sqrt{2}
   \sqrt[4]{\pi } \sqrt[4]{{g_s}} \sqrt[4]{N} r}{\sqrt{r^4-{r_h}^4}} +  \frac{27 \left(9 b^2+1\right)^4 \beta  b^{10} \sqrt[4]{{g_s}} M r^2 \Sigma \left(6 a^2+{r_h}^2\right)
   (r-2 {r_h}) \log ^3({r_h})}{2 \sqrt{2} \pi ^{3/4} \left(3 b^2-1\right)^5 \left(6 b^2+1\right)^4 \log N ^4
   N {N_f} {r_h}^4 \alpha _{\theta _2}^3 \left(9 a^2+{r_h}^2\right) \sqrt{r^4-{r_h}^4}},\nonumber
\end{eqnarray}   
   \begin{eqnarray}
   & & \hskip -0.4in \omega_r^{\underline{8}\ \underline{10}} =  -\frac{7
   N^{3/5}}{{\lambda_5} {g_s}^{3/2} \kappa_2
   \log N  {N_f} r^2 \alpha _{\theta _1}^4
   \alpha _{\theta _2} \left(r^2-3.3 a^2\right)}\nonumber\\
& & \hskip -0.4in  + \frac{\kappa_{\omega_r^{\underline{8}\ \underline{10}}} a^8
   \sqrt[4]{\beta } \sqrt{{\cal C}_{zz}}
   {\rm const} {\lambda_5} {g_s}^{5/4} M N^{19/20}
   {N_f} \sqrt{\alpha _{\theta _1}}
   \sqrt{1-\frac{{r_h}^4}{r^4}} \log (r)}{r^4
   \alpha _{\theta _2}^6 \left(r^2-3. a^2\right)^2
   \sqrt{\frac{6 a^2+r^2}{9 a^2+r^2}}},\end{eqnarray}
}

\noindent with $\Sigma \equiv -19683 \sqrt{6} \alpha _{\theta _1}^6-6642
   \alpha _{\theta _2}^2 \alpha _{\theta _1}^3+40 \sqrt{6} \alpha _{\theta _2}^4$, and setting consistently the ${\cal O}(R^4)$ corrections of ${\cal M}$-theory's three-form potential to zero requires: ${\cal C}_{zz} - 2 {\cal C}_{\theta_1z}=0$ and $|{\cal C}_{\theta_1x}|\ll1$ \cite{OR4}, we see that $E^r_{\underline{5}}$
receives no ${\cal O}(\beta)$ corrections. Further, the constants appearing in the EOM for massive mesinos are therefore given below:
{\footnotesize
\begin{eqnarray}
\label{constants-exps}
& & {\cal C}_{\frac{E^t_{\underline{1}}}{E^r_{\underline{5}}}}^{M_{\rm Mesino}} = \frac{\beta  \sqrt{{g_s}} M \left(19683 \sqrt{6} \alpha _{\theta _1}^6+6642 \alpha _{\theta _2}^2 \alpha _{\theta
   _1}^3-40 \sqrt{6} \alpha _{\theta _2}^4\right) \log ^3({r_h})}{17496 \sqrt{\pi } \epsilon ^5 \left(\log N\right)^4
   N^{3/4} {N_f} {r_h}^3 \alpha _{\theta _2}^3}+\frac{\sqrt{3 \pi } \sqrt{{g_s}} \sqrt{N}}{4 {r_h}},\nonumber\\
   & & {\cal C}_{\left(\frac{E^r_{\underline{5}}}{E^t_{\underline{1}}}\right)'}^{M_{\rm Mesino}} =
   \frac{2 \beta  M \left(-19683 \sqrt{6} \alpha _{\theta _1}^6-6642 \alpha _{\theta _2}^2 \alpha _{\theta _1}^3+40
   \sqrt{6} \alpha _{\theta _2}^4\right) \log ^3({r_h})}{6561 \pi ^{3/2} \epsilon ^5 \sqrt{{g_s}} \left(\log N\right)^4
   N^{7/4} {N_f} {r_h} \alpha _{\theta _2}^3}+\frac{4 {r_h}}{\sqrt{3 \pi } \sqrt{{g_s}} \sqrt{N}},\nonumber\\
  & &  {\cal C}_{rt}^{M_{\rm Mesino}} = -\frac{14}{3 \sqrt{3 \pi } \sqrt{{g_s}} \sqrt{N}},\nonumber\\
  & & {\cal C}_{\frac{E^r_{\underline{5}}}{E^t_{\underline{1}}}}^{M_{\rm Mesino}} = \frac{4 {r_h}}{\sqrt{3 \pi } \sqrt{{g_s}} \sqrt{N}}+\frac{4 \beta  M \left(-19683 \sqrt{6} \alpha _{\theta _1}^6-6642 \alpha _{\theta _2}^2 \alpha _{\theta _1}^3+40
   \sqrt{6} \alpha _{\theta _2}^4\right) \sqrt{\frac{6 a^2+{r_h}^2}{9 a^2+{r_h}^2}} \log ^3({r_h})}{6561
   \sqrt{3} \pi ^{3/2} \epsilon ^5 \sqrt{{g_s}} \left(\log N\right)^4 N^{7/4} {N_f} {r_h} \alpha _{\theta
   _2}^3},\nonumber\\
& & {\cal C}_{\frac{E^{x^1}_{\underline{2}}}{E^t_{\underline{1}}}}^{M_{\rm Mesino}}
= \frac{2 \sqrt{2} \sqrt[4]{\pi } \sqrt[4]{{g_s}} \sqrt[4]{N}}{\sqrt{{r_h}^3}}-\frac{4 \sqrt{\frac{2}{3}} \beta 
   \sqrt[4]{{g_s}} M \left(-19683 \sqrt{6} \alpha _{\theta _1}^6-6642 \alpha _{\theta _2}^2 \alpha _{\theta _1}^3+40
   \sqrt{6} \alpha _{\theta _2}^4\right) \left(6 a^2+{r_h}^2\right) \log ^3({r_h})}{6561 \pi ^{3/4} \epsilon ^5
   \left(\log N\right)^4 N {N_f} {r_h}^{7/2} \alpha _{\theta _2}^3 \left(9 a^2+{r_h}^2\right)},\nonumber\\
 & & {\cal C}_{\frac{E^{x^1}_{\underline{2}}}{E^r_{\underline{5}}}}^{M_{\rm Mesino}} =
\frac{\sqrt{\frac{3}{2}} \pi ^{3/4} {g_s}^{3/4} N^{3/4}}{{r_h}^{3/2}}
 + \frac{\beta  {g_s}^{3/4} M \sqrt{\frac{1}{N}} \left(19683 \sqrt{6} \alpha _{\theta _1}^6+6642 \alpha _{\theta _2}^2
 \alpha _{\theta _1}^3-40 \sqrt{6} \alpha _{\theta _2}^4\right) \log ^3({r_h})}{2187 \sqrt{2} \sqrt[4]{\pi } \epsilon ^5 \left(\log N\right)^4 {N_f} {r_h}^{7/2} \alpha _{\theta _2}^3}
,\nonumber\\
   & & a_{2\ \beta}^{M_{\rm Mesino}} \equiv -\frac{21.9}{\sqrt{g_s N}} + {\cal C}_{rt}^{M_{\rm Mesino}}\nonumber\\
   & & {\cal A}_1^{M_{\rm Mesino}} = \pi ^2 \left(-(2 n+1)^2\right) T^2 \Biggl(\frac{\beta  \sqrt{{g_s}} M \left(19683 \sqrt{6} \alpha _{\theta _1}^6+6642
   \alpha _{\theta _2}^2 \alpha _{\theta _1}^3-40 \sqrt{6} \alpha _{\theta _2}^4\right) \log ^3({r_h})}{17496
   \sqrt{\pi } \epsilon ^5 \left(\log N\right)^4 N^{3/4} {N_f} {r_h}^3 \alpha _{\theta _2}^3}\nonumber\\
   & & +\frac{\sqrt{3 \pi }
   \sqrt{{g_s}} \sqrt{N}}{4 {r_h}}\Biggr)-(0.\, +3.3 i) (2 n+1) T,\nonumber\\
   & & {\cal A}_2^{M_{\rm Mesino}} = \frac{2 \beta  M \left(-19683 \sqrt{6} \alpha _{\theta _1}^6-6642 \alpha _{\theta _2}^2 \alpha _{\theta _1}^3+40
   \sqrt{6} \alpha _{\theta _2}^4\right) \log ^3({r_h})}{6561 \pi ^{3/2} \epsilon ^5 \sqrt{{g_s}} \left(\log N\right)^4
   N^{7/4} {N_f} {r_h} \alpha _{\theta _2}^3}-\frac{1.3 {r_h}}{\sqrt{{g_s}} \sqrt{N}}+\frac{4
   {r_h}}{\sqrt{3 \pi } \sqrt{{g_s}} \sqrt{N}}+2 i \pi  (2 n+1) T,\nonumber\\
   & & {\cal B}_1^{M_{\rm Mesino}} = 1.1 i {\cal C}_{\frac{E^{x^1}_{\underline{2}}}{E^t_{\underline{1}}}}^{M_{\rm Mesino}} + (2n+1)T {\cal C}_{\frac{E^{x^1}_{\underline{2}}}{E^r_{\underline{5}}}}^{M_{\rm Mesino}}. 
   \end{eqnarray}
   }

\end{itemize}

\section{$\tilde{z}=$Constant Embedding of Flavor $D6$-Branes Inclusive of ${\cal O}(\beta)$ Corrections}
\setcounter{equation}{0}
\seceqcc
\label{ztilde-const-incl-beta}

The EOM for the embedding of the flavor $D6$-branes in the warped squashed resolved conifold $\tilde{z} = \tilde{z}(r) = \tilde{z}_{(0)} + \beta \tilde{z}_{(1)}$, up to ${\cal O}(\beta)$, is given by:
{\footnotesize
\begin{eqnarray}
\label{ztilde-EOM-IR-i}
& & \hskip -0.5in \frac{N^{3/5} {N_f} r^2 \left(r^4-{r_h}^4\right) (\log N -3 \log ({r_h})) \left(\tilde{z}_{(0)}'+\beta 
   \tilde{z}_{(1)}'\right)}{4 \sqrt{6} \pi ^{7/4} {g_s}^{3/4} \alpha _{\theta _1}^2 \alpha _{\theta _2}^{5/2}
   \sqrt{\frac{4 \sqrt{\pi } \sqrt{{g_s}} r^2 \alpha _{\theta _2}^3 \left(6 a^2+r^2\right)}{9 a^2+r^2}+3 \sqrt{3}
   N^{2/5} \left(r^4-{r_h}^4\right) \left(\tilde{z}_{(0)}'+\beta  \tilde{z}_{(1)}'\right)^2}}\nonumber\\
& & \hskip -0.5in -\frac{0.0005 \beta  M
   N^{87/20} {r_h}^5 \left(-492.1 \alpha _{\theta _1}^6-67.8 \alpha _{\theta _2}^2 \alpha _{\theta _1}^3+\alpha
   _{\theta _2}^4\right) (r-{r_h})^2 \log ^3({r_h}) (\log N -3 \log ({r_h}))
   \tilde{z}_{(0)}'(r)}{\epsilon ^5 {g_s}^{5/2} \log N ^4 \alpha _{\theta _1}^6 \alpha _{\theta _2}^6 \left(9
   a^2+{r_h}^2\right)} = {K^{(0)}}+\beta  {K^{(1)}}.
\end{eqnarray}
}
At ${\cal O}(\beta^0)$, (\ref{ztilde-EOM-IR-i}) yields:
{\footnotesize
\begin{eqnarray}
\label{ztilde-EOM-IR-ii}
& & \tilde{z}_{(0)}' =\pm \frac{8 \sqrt{6} \pi ^2 {g_s}^{3/4} {K^{(0)}} \alpha _{\theta _1}^2 \alpha _{\theta _2}^{5/2}
   \sqrt{\frac{\sqrt{{g_s}} r^2 \alpha _{\theta _2}^3 \left(6 a^2+r^2\right)}{9 a^2+r^2}}}{\sqrt{N^{2/5}
   \left(r^4-{r_h}^4\right) \left(N^{4/5} {N_f}^2 r^4 \left(r^4-{r_h}^4\right) (\log N -3 \log
   ({r_h}))^2-288 \sqrt{3} \pi ^{7/2} {g_s}^{3/2} {K^{(0)}}^2 \alpha _{\theta _1}^4 \alpha _{\theta
   _2}^5\right)}}.
\end{eqnarray}
}
From (\ref{ztilde-EOM-IR-ii}), one obtains $\tilde{z}(r)\in\mathbb{R}$ if ${K^{(0)}}=0$ (irrespective of whether one performs first a large-$N$ followed by a small-$r$ expansion or vice versa). In a similar manner, at ${\cal O}(\beta)$,
{\footnotesize 
\begin{eqnarray}
\label{ztilde-EOM-IR-iii}
& & \hskip -0.4in \tilde{z}_{(1)} = c_1 \nonumber\\
& & \hskip -0.4in -\frac{4 \pi ^2 {g_s}^{5/4} {K^{(1)}} r \alpha _{\theta _1}^2 \alpha _{\theta _2}^{11/2} \sqrt{6 a^2+r^2}
  }{N^{3/5} {N_f} {r_h}^4
   \sqrt{9 a^2+r^2} \sqrt{9 a^2-{r_h}^2} \sqrt{9 a^2+{r_h}^2} (\log N -3 \log ({r_h}))
   \sqrt{\frac{\sqrt{{g_s}} r^2 \alpha _{\theta _2}^3 \left(6 a^2+r^2\right)}{9 a^2+r^2}}}\nonumber\\
   & & \hskip -0.4in \times  \Biggl(\sqrt{6} \left(\sqrt{{r_h}^2-6 a^2} \sqrt{9 a^2+{r_h}^2} \tan ^{-1}\left(\frac{\sqrt{6 a^2+r^2} \sqrt{9
   a^2-{r_h}^2}}{\sqrt{9 a^2+r^2} \sqrt{{r_h}^2-6 a^2}}\right)+\sqrt{-6 a^2-{r_h}^2} \sqrt{9
   a^2-{r_h}^2} \tan ^{-1}\left(\frac{\sqrt{6 a^2+r^2} \sqrt{9 a^2+{r_h}^2}}{\sqrt{9 a^2+r^2} \sqrt{-6
   a^2-{r_h}^2}}\right)\right)\nonumber\\
& & \hskip -0.4in   -4 \sqrt{9 a^2-{r_h}^2} \sqrt{9 a^2+{r_h}^2} \tanh
   ^{-1}\left(\frac{\sqrt{\frac{3}{2}} \sqrt{6 a^2+r^2}}{\sqrt{9 a^2+r^2}}\right)\Biggr).\nonumber\\
& &    
\end{eqnarray}
}
Again, a finite real $\tilde{z}_{(1)}$ is obtained only for ${K^{(1)}}=0$. This hence implies a constant $\tilde{z}$-embedding up to ${\cal O}(\beta)$.

\section{Constants appearing in the Solution to the Mesino Wave-Function for $T>T_c$}
\label{gy}
\setcounter{equation}{0}
\seceqdd

The parameters $\mu_{1, 2, 3}, \Lambda$ in equation (\ref{solution-high-T}), are defined as follows (in the following, terms of ${\cal O}\left(\frac{\beta}{N^\alpha}\right),\ \alpha\geq1$ have been dropped as the same were sub-dominant as compared to the order considered in the \cite{OR4}):   
{\footnotesize
\begin{eqnarray*}
& & \mu_1 \equiv -\frac{{\cal A}_2^{M_{\rm Mesino}}}{{{\cal C}_{\frac{E^r_{\underline{5}}}{E^t_{\underline{1}}}}^{M_{\rm Mesino}}}\ }+\frac{2 i
   {{\cal B}_1^{M_{\rm Mesino}}}}{{{\cal C}_{\frac{E^{x^1}_{\underline{2}}}{E^t_{\underline{1}}}}^{M_{\rm Mesino}}}}+\frac{\sqrt{{\cal A}_2^{M_{\rm Mesino}}\ ^2-2 {{\cal C}_{\frac{E^r_{\underline{5}}}{E^t_{\underline{1}}}}^{M_{\rm Mesino}}}\ 
   {\cal A}_2^{M_{\rm Mesino}}+{{\cal C}_{\frac{E^r_{\underline{5}}}{E^t_{\underline{1}}}}^{M_{\rm Mesino}}}\ ^2+4 {{\cal A}_1^{M_{\rm Mesino}}}\  {{\cal C}_{\frac{E^r_{\underline{5}}}{E^t_{\underline{1}}}}^{M_{\rm Mesino}}}\ }}{{{\cal C}_{\frac{E^r_{\underline{5}}}{E^t_{\underline{1}}}}^{M_{\rm Mesino}}}\ }+1\nonumber\\
& & = 2.5 \sqrt{-(0.3\, +0.4 i) n+(0.1\, -0.2 i)-0.3 n^2}+(0.9\, -0.9 i) n-(0.7\, +0.4 i)-\frac{0.6 (n+0.5)
   {r_h}}{\sqrt{{g_s}} \sqrt{N}} \nonumber\\
   & & + {\cal O}\left(\cancelto{0}{\frac{\beta}{N^{5/4}}}\right),\nonumber\\   
\end{eqnarray*}
\begin{eqnarray} 
\label{arguments-U}
& & \mu_2 \equiv \frac{2
   \sqrt{{\cal A}_2^{M_{\rm Mesino}}\ ^2-2 {{\cal C}_{\frac{E^r_{\underline{5}}}{E^t_{\underline{1}}}}^{M_{\rm Mesino}}}\  {\cal A}_2^{M_{\rm Mesino}}+{{\cal C}_{\frac{E^r_{\underline{5}}}{E^t_{\underline{1}}}}^{M_{\rm Mesino}}}\ ^2+4 {{\cal A}_1^{M_{\rm Mesino}}}\ 
   {{\cal C}_{\frac{E^r_{\underline{5}}}{E^t_{\underline{1}}}}^{M_{\rm Mesino}}}\ }}{{{\cal C}_{\frac{E^r_{\underline{5}}}{E^t_{\underline{1}}}}^{M_{\rm Mesino}}}\ }+1\nonumber\\
& & = 1+4.9 \sqrt{-(0.25\, +0.4 i) n+(0.1\, -0.2 i)-0.3 n^2} + {\cal O}\left(\cancelto{0}{\frac{\beta}{N^{5/4}}}\right),\nonumber\\ 
& &  \mu_3 \equiv  \frac{i \sqrt{6} \pi ^{3/4}
   {g_s}^{3/4} N^{3/4} {p} u}{{r_h}^{5/2}} +  \frac{2 i {{\cal C}_{\frac{E^{x^1}_{\underline{2}}}{E^t_{\underline{1}}}}^{M_{\rm Mesino}}} {p}}{{{\cal C}_{\frac{E^r_{\underline{5}}}{E^t_{\underline{1}}}}^{M_{\rm Mesino}}}\ }   \nonumber\\
  & & =  \frac{i \sqrt{6} \pi ^{3/4}
   {g_s}^{3/4} N^{3/4} {p} u}{{r_h}^{5/2}} + \frac{i \sqrt{2} \beta  {g_s}^{3/4} M {p} u \left(19683 \sqrt{6} \alpha _{\theta _1}^6+6642 \alpha _{\theta
   _2}^2 \alpha _{\theta _1}^3-40 \sqrt{6} \alpha _{\theta _2}^4\right) \log ^3({r_h})}{2187 \sqrt[4]{\pi } \epsilon
   ^5 \log N ^4 \sqrt{N} {N_f} {r_h}^{9/2} \alpha _{\theta _2}^3}\nonumber\\
& & = \frac{i \sqrt{6} \pi ^{3/4}{g_s}^{3/4} N^{3/4} {p} u}{{r_h}^{5/2}}  
+ {\cal O}\left(\cancelto{0}{\frac{\beta}{N}}\right);\nonumber\\
& & \Lambda \equiv \mu_1 - \frac{2 i
   {{\cal B}_1^{M_{\rm Mesino}}}}{{{\cal C}_{\frac{E^{x^1}_{\underline{2}}}{E^t_{\underline{1}}}}^{M_{\rm Mesino}}}}= {2.5 \sqrt{-(0.3\, +0.4 i) n+(0.1\, -0.2 i)-0.3 n^2}+(-0.9 i) n+(1\, -0.4 i) + {\cal O}\left(\cancelto{0}{\frac{\beta}{N^{5/4}}}\right)}.\nonumber\\
& &    
\end{eqnarray}
}

\section{Summary of Applications of Top-Down Holographic QCD \cite{MQGP}, \cite{OR4}}
\label{Applications_MQGP}
\setcounter{equation}{0}
\seceqee

One of the authors (AM) has been working on the top-down holographic QCD for the past few years. The holographic dual of finite $N$ QCD was first constructed in \cite{MQGP} and then ${\cal O}(R^4)$ corrections to \cite{MQGP} were obtained in \cite{OR4}. Following is the summary of results obtained in this direction.

\begin{itemize}

\item {\bf Summary of Applications of \cite{MQGP}}:
In \cite{DM-transport-2014}, transport coefficients such as shear
viscosity, diffusion constant, electrical conductivity, charge susceptibility, etc., of black $M3$-banes (black $M5$-branes wrapping a homologous sum of two cycles) in the MQGP limit were obtained, and it was found that the ratio of shear viscosity-to-entropy density is $1/4\pi$. In \cite{NPB}, deconfinement temperature and mass scale of the first generation quarks were obtained without the inclusion of ${\cal O}(R^4)$ corrections relevant to thermal QCD. Further, thermodynamic stability and $G_2$ structure of \cite{MQGP} and temperature dependence of electrical conductivity and charge susceptibility were also discussed in \cite{NPB}, \cite{MG14}. In this process, Einstein’s law was verified by computing the ratio of electrical conductivity to charge susceptibility. For the discussion on Wiedemann-Franz law by calculating the thermal and electrical conductivities up to LO in $N$ and NLO in $N$ correction to the aforementioned transport coefficients and speed of sound from the gauge invariant metric perturbations, see \cite{K+A}. The glueball and meson spectra of finite $N$ QCD have been obtained in \cite{IITR-glueballs} and \cite{V+A+K}, respectively. Decay of glueballs into mesons ($\pi$ and $\rho$ mesons) has been discussed in \cite{VA-Glueball-decay} and for the QCD trace anomaly from ${\cal M}$-theory perspective, see \cite{Trace-Anomaly}.

\item {\bf Summary of Applications of \cite{OR4}}: The low energy coupling constants at the NLO in chiral expansion of $SU(3)$ chiral perturbation theory (for simplicity in the chiral limit) were obtained from the aforementioned type IIA dual, in \cite{Vikas+Gopal+Aalok} where we observed a {\it connection between higher derivative terms and large-$N$ expansion}. In the process of computing the deconfinement temperature $(T_c)$ in \cite{Gopal+Vikas+Aalok,Gopal-Tc-Vorticity}, a {\it novel ``UV-IR'' mixing, non-renormalization $T_c$ beyond one-loop in the zero instanton sector and ``Flavor Memory'' effect} were obtained. Further, we constructed a doubly holographic setup with a non-conformal bath in \cite{Gopal+Aalok} to get the Page curve of the related eternal black hole from a top-down approach. One of the exciting results that we obtained in \cite{Gopal+Aalok} is the Page curve of the relevant eternal black hole for massless gravity on the Karch-Randall brane. Massless graviton was responsible for the exponential-in-$N$ suppressed entanglement entropy from higher derivative terms in eleven-dimensional supergravity action. This provided us the {\it connection between the mass of graviton and higher derivative terms}. On the Math side with the aim of classifying non-supersymmetric thermal geometries relevant to realistic top-down holographic duals of thermal QCD-like theories,  $SU(3)/G_2/SU(4)/Spin(7)$-structures and (Almost) Contact (3) (Metric) Structures on the underlying six-, seven- and eight-folds were studied in \cite{OR4} and \cite{ACMS}.

\end{itemize}


\begin{thebibliography}{99}
\bibitem{MQGP} M.~Dhuria and A.~Misra, {\it Towards MQGP}, JHEP 1311 (2013) 001 [arXiv:hep-th/1306.4339].
\bibitem{metrics} M.~Mia, K.~Dasgupta, C.~Gale and S.~Jeon, {\it Five Easy Pieces: The Dynamics of Quarks in Strongly Coupled Plasmas}, Nucl.\ Phys.\ B {\bf 839}, 187 (2010) [arXiv:hep-th/0902.1540].
\bibitem{OR4} A.~Misra and V.~Yadav, {\it On ${\cal M}$-Theory Dual of Large-$N$ Thermal QCD-Like Theories up to ${\cal O}(R^4)$ and $G$-Structure Classification of Underlying Non-Supersymmetric Geometries}, to appear in Advances in Theoretical and Mathematical Physics (2022) {\bf 26}, No. 10 [arXiv:2004.07259 [hep-th]].
\bibitem{Maldacena} J.~Maldacena, {\it The Large N Limit of Superconformal Field Theories and Supergravity}, Adv.Theor.Math.Phys.2:231-252,1998 [arXiv:hep-th/9711200].
\bibitem{Vikas+Gopal+Aalok} V.~Yadav, G.~Yadav and A.~Misra,{\it (Phenomenology/Lattice-Compatible) $SU(3)$ M$\chi$PT HD up to ${\cal O}(p^4)$ and the ${\cal O}\left(R^4\right)$-Large-$N$ Connection}, JHEP 2108 (2021) 151 [arXiv:2011.04660 [hep-th]].
\bibitem{Fermionic-action} D.~Marolf, L.~Martucci and P. J.~Silva, {\it Fermions, T-duality and effective actions for
D-branes in bosonic backgrounds}, JHEP {\bf 0304} (2003) 051 [arXiv:hep-th/0303209]; D.~Marolf, L.~Martucci and P. J.~Silva, {\it Actions and fermionic symmetries for D-branes in bosonic backgrounds}, JHEP {\bf 0307} (2003) 019  [arXiv:hep-th/0306066]; L.~Martucci and J.Rosseel, {\it Dirac actions for D-branes
on backgrounds with fluxes}, Class.Quant.Grav. 22 (2005) 2745-2764 [arXiv:hep-th/050404].
\bibitem{Abidin+Carlson} Z.~Abidin, Carl~E.~Carlson, {\it Nucleon electromagnetic and gravitational form factors from holography}, 	Phys. Rev. D {\bf 79} (2009) 115003 [arXiv:0903.4818[hep-ph]].
\bibitem{SS} T.~Sakai, S.~Sugimoto, {\it Low energy hadron physics in holographic QCD}, Prog.Theor.Phys.113:843-882,2005 [arXiv:hep-th/0412141]; T.~Sakai, S.~Sugimoto, {\it More on a holographic dual of QCD}, Prog.Theor.Phys.114:1083-1118,2005 [arXiv:hep-th/0507073].
\bibitem{mesino-SS} R.~Heise and  H.~G.~Svendsen, {\it A note on fermions in holographic QCD}, JHEP {\bf 08} (2007) 065 [arXiv:0706.2253[hep-th]].
\bibitem{ACMS} A.~Misra and G.~Yadav, {\it (Almost) Contact (3) (Metric) Structure(s) and Transverse $SU(3)$ Structures Associated with ${\cal M}$-Theory Dual of Thermal QCD at Intermediate Coupling }, [arXiv:2211.13186[hep-th]].
\bibitem{M.Ionel and M.Min-OO (2008)} M.~Ionel and M.~Min-Oo, {\it Cohomogeneity One Special Lagrangian 3-Folds in the Deformed and the Resolved Conifolds},  Illinois Journal of Mathematics, Vol. 52, Number 3 (2008).	
\bibitem{NPB} K.~Sil and A.~Misra, {\it On Aspects of Holographic Thermal QCD at Finite Coupling},
  Nucl.\ Phys.\ B {\bf 910}, 754 (2016) [arXiv:1507.02692 [hep-th]].
\bibitem{VA-Glueball-decay} V.~Yadav and A.~Misra, {\it M-Theory Exotic Scalar Glueball Decays to Mesons at Finite Coupling}, JHEP {\bf 09}, 133 (2018) [arXiv:1808.01182 [hep-th]].
\bibitem{Kruczenski et al-2003} M.~Kruczenski, D.~Mateos, R.~C.~Myers and D.~J.~Winters, {\it Towards a holographic dual of large-$N_c$ QCD}, JHEP 0405 (2004) 041 [arXiv:hep-th/0311270].
\bibitem{DM-transport-2014} M.~Dhuria and A.~Misra, {\it Transport Coefficients of Black MQGP M3-Branes}, Eur. Phys. J. C {\bf 75}, 16 (2015) [arXiv:1406.6076[hep-th]].
\bibitem{Armoni et al-2020} R.~Argurio, A.~Armoni, M.~Bertolini, F.~Mignosa and P.~Niro, {\it Vacuum structure of large N QCD$_3$ from holography}, JHEP {\bf 07}, 134 (2020) [arXiv:2006.01755[hep-th]].
\bibitem{Becker-sisters-O(R^4)}K.~Becker and M.~Becker, {\it Supersymmetry Breaking, M-Theory and Fluxes}, JHEP {\bf 07}, 038 (2001) [arXiv:hep-th/0107044].
\bibitem{Nakas+Rigatos}T.~Nakas and K.~S.~Rigatos,
{\it Fermions and baryons as open-string states from brane junctions},
JHEP \textbf{12}, 157 (2020) [arXiv:2010.00025 [hep-th]].
\bibitem{V+A+K} V.~Yadav, A.~Misra and K.~Sil, {\it Delocalized SYZ Mirrors and Confronting Top-Down $SU(3)$-Structure Holographic Meson Masses at Finite $g$ and $N_c$ with P(article) D(ata) G(roup) Values}, Eur.Phys.J.C 77 (2017) 10, 656 [arXiv:1707.02818[hep-th]].
%
\bibitem{Witten-Hawking-Page-Tc} E.~Witten, {\it Anti-de Sitter space, thermal phase transition, and confinement in gauge theories}, Adv. Theor. Math. Phys. {\bf 2}, 505 (1998) [arXiv:hep-th/9803131].
\bibitem{Green and Gutperle and Vanhove} M.~B.~Green and M.~Gutperle, {\it Effects of D-instantons}, Nucl. Phys. B {\bf 498}, 195-227 (1997) [arXiv:hep-th/9701093]; M.~B.~Green and P.~Vanhove, {\it D-instantons, Strings and M-theory}, Phys. Lett. B {\bf 408}, 122-134 (1997) [arXiv:hep-th/9704145].
\bibitem{IITR-glueballs} K.~Sil, V.~Yadav and A.~Misra,
{\it Top-down holographic G-structure glueball spectroscopy at (N)LO in $N$ and finite coupling}, Eur. Phys. J. C {\bf 77}, no.6, 381 (2017) [arXiv:1703.01306 [hep-th]].
\bibitem{Bulk-Viscosity-McGill-IIT-Roorkee} A.~Czajka, K.~Dasgupta, C.~Gale, S.~Jeon, A.~Misra, M.~Richard and K.~Sil,
{\it Bulk Viscosity at Extreme Limits: From Kinetic Theory to Strings}, JHEP {\bf 07}, 145 (2019)[arXiv:1807.04713 [hep-th]].
\bibitem{Kirsch} I.~Kirsch, {\it Spectroscopy of fermionic operators in AdS/CFT}, JHEP {\bf 09}, 052 (2006) [arXiv:hep-th/0607205].
\bibitem{Gopal+Vikas+Aalok} G.~Yadav, V.~Yadav and A.~Misra, {\it $\mathscr {M}$cTEQ ($\mathscr {M}$ chiral perturbation theory-compatible deconfinement Temperature and Entanglement entropy up to terms Quartic in curvature) and FM (Flavor Memory)},  JHEP {\bf 10} (2021) 220, arXiv:2108.05372 [hep-th].
\bibitem{Witten-chiral-cond_mq} E.~Witten, {\it Multi-Trace Operators, Boundary Conditions, And AdS/CFT Correspondence}, arXiv:hep-th/0112258 [hep-th].
\bibitem{mqqbarq-RGI} J. C. Collins, {\it Renormalization}, ~Cambridge University Press, Cambridge, England, 1984; J.~L.~Kneur,
{\it Variational quark mass expansion and the order parameters of chiral symmetry breaking}, Phys. Rev. D \textbf{57}, 2785-2805 (1998) [arXiv:hep-ph/9609265  [hep-ph]].
\bibitem{Gopal+Aalok} G.~Yadav and A.~Misra, {\it Entanglement entropy and Page curve from the M-theory dual of thermal QCD above  $T_c$ at intermediate coupling}, Phys. Rev. D {\bf 107} (2023) 10, 106015 [arXiv:2207.04048[hep-th]].
\bibitem{MG14} A.~Misra, {\it Resolved warped deformed conifolds and large-N thermal QCD via black M3-branes}, Contribution to MG14, 4211-4215.
\bibitem{K+A} K.~Sil and A.~Misra, {\it New Insights into Properties of Large-N Holographic Thermal QCD at Finite Gauge Coupling at (the Non-Conformal/Next-to) Leading Order in N}, Eur.Phys.J. C 76 (2016) no.11, 618 [arXiv:1606.04949 [hep-th]].
\bibitem{Trace-Anomaly} A.~Misra and C.~Gale, {\it The QCD trace anomaly at strong coupling from  M-theory }, Eur. Phys. J. C bf{80}, no.7, 620 (2020)
[arXiv:1909.04062 [hep-th]].
\bibitem{Gopal-Tc-Vorticity} G.~Yadav, {\it Deconfinement temperature of rotating QGP at intermediate coupling from 
M-theory}, Phys. Lett. B 841 (2023) 137925 [arXiv:2203.11959 [hep-th]]

\end{thebibliography}
\end{document}